\documentclass[lettersize,journal]{IEEEtran}
\usepackage{amsmath,amsfonts}
\usepackage{subfigure}
\usepackage{algorithm}
\usepackage{array}
\usepackage[caption=false,font=normalsize,labelfont=sf,textfont=sf]{subfig}
\usepackage{textcomp}
\usepackage{stfloats}
\usepackage{url}
\usepackage{verbatim}
\usepackage{graphicx}
\usepackage{cite}
\usepackage{comment}
\usepackage[acronym]{glossaries}
\usepackage{xcolor}  

\usepackage{algpseudocode}
\usepackage[inline,shortlabels]{enumitem}

\newcommand\Algphase[1]{%
\vspace*{-.7\baselineskip}\Statex\hspace*{\dimexpr-\algorithmicindent-2pt\relax}\hrule%
\Statex\hspace*{-\algorithmicindent}\textbf{#1}%
\vspace*{-.7\baselineskip}\Statex\hspace*{\dimexpr-\algorithmicindent-2pt\relax}\hrule%
}

\newcommand{\ie}[1]{\textit{i.e.,}}
\newcommand{\eg}[1]{\textit{e.g.,}}

\begin{document}

\title{{\normalsize This work has been submitted to the IEEE for possible publication in Transactions on Network and Service\vspace{-0.6cm} Management~(TNSM). Copyright may be transferred without notice, after which this version may no longer be accessible.\vspace{0.8cm}}
Measurement Study of Programmable Network Coding in Cloud-native 5G and Beyond Networks
\thanks{Abridged preliminary summaries of this work appeared
    in~\cite{osel2023}.}}

\author{\IEEEauthorblockN{
Osel Lhamo \IEEEauthorrefmark{1},
Tung V. Doan \IEEEauthorrefmark{2},
Elif Tasdemir \IEEEauthorrefmark{1},
Mahdi Attawna \IEEEauthorrefmark{1},
Giang T. Nguyen \IEEEauthorrefmark{2}\IEEEauthorrefmark{3},\\
Patrick Seeling \IEEEauthorrefmark{4},
Martin Reisslein \IEEEauthorrefmark{5},
Frank H.P. Fitzek\IEEEauthorrefmark{1}\IEEEauthorrefmark{3} \\
}
\IEEEauthorblockA{\textit{\IEEEauthorrefmark{1}Deutsche Telekom Chair of Communication Networks, TU Dresden, Germany} \\
\textit{\IEEEauthorrefmark{2}Haptic Communication Systems, TU Dresden, Germany}\\
\textit{\IEEEauthorrefmark{3}Centre for Tactile Internet with Human-in-the-Loop (CeTI)}\\
\textit{\IEEEauthorrefmark{4}Department of Computer Science, Central Michigan University}\\
\textit{\IEEEauthorrefmark{5}School of Electrical, Computer, and Energy Engineering, Arizona State University}}
}

\maketitle

\begin{abstract}
Emerging 5G/6G use cases span various industries, necessitating flexible solutions that leverage emerging technologies to meet diverse and stringent application requirements under changing network conditions. The standard 5G RAN solution, retransmission, reduces packet loss but can increase transmission delay in the process. Random Linear Network Coding (RLNC) offers an alternative by proactively sending combinations of original packets, thus reducing both delay and packet loss. Current research often only simulates the integration of RLNC in 5G while we implement and evaluate our approach on real commercially available hardware in a real-world deployment. 

We introduce Flexible Network Coding (FlexNC), which enables the flexible fusion of several RLNC protocols by incorporating a forwarder with multiple RLNC nodes. Network operators can configure FlexNC based on network conditions and application requirements. To further boost network programmability, our Recoder in the Network (RecNet) leverages intermediate network nodes to join the coding process.
Both the proposed algorithms have been implemented on OpenAirInterface and extensively tested with traffic from different applications in a real network. While FlexNC adapts to various application needs of latency and packet loss, RecNet significantly minimizes packet loss for a remote user with minimal increase in delay compared to pure RLNC.

\end{abstract}


\section{Introduction}

The telecommunications industry has rapidly transitioned to cloud-based mobile networks. Traditionally, Mobile Cloud Computing (MCC) enabled users to access cloud services via the Internet. With the advent of low-latency applications such as the Tactile Internet (TI), the Mobile Edge Cloud concept emerged, bringing cloud services closer to users at base stations. Recently, mobile operators have integrated cloud computing into mobile core and Radio Access Networks (RANs), forming cloud-native 5G and beyond networks. This shift promises significant benefits, including cost reduction and faster time-to-market. 
Ultra Reliable and Low Latency Communication (URLLC) has become a key focus due to its importance for applications such as network-assisted driving and remote surgery. For example, packet loss in network-assisted driving could result in unexpected behaviours and accidents, highlighting the critical need for reliability.  However, transforming existing mobile core and RAN components into cloud-native systems presents challenges in meeting these stringent requirements. 

Achieving high application reliability in cloud-native 5G and beyond networks requires \textit{flexible solutions}, due to, e.g., 
\begin{enumerate*}[i.)]
    \item highly diverse application requirements, e.g., for latency or throughput, making fixed solutions ill-suited;
    \item applications with varying reliability requirements, e.g., some can tolerate a few lost packets, while others need five-nines reliability;
    \item changing network conditions, such as lossy links and congestion; or
    \item emerging technologies like Network Function Virtualization (NFV).
\end{enumerate*}
Much effort, ranging from standardization to research studies, has been made to study the reliability of mobile networks. The 3GPP, focusing on RANs and simplicity, has proposed retransmission mechanisms between gNodeB (gNB) and User Equipment (UE). While effective in reducing packet loss, this method introduces significant delays especially under lossy conditions, making it unsuitable for low-latency applications. 
Research on improving the reliability of mobile networks has primarily focused on the physical layer, using techniques such as coverage enhancement~\cite{8740799,9330587} or new radio access technologies~\cite{8723481}. However, these methods require changes to the existing RAN architecture, limiting their adoption. Alternatively, studies have proposed using network coding~\cite{vieira2017network,torres2015network,7263352}, specifically Random Linear Network Coding (RLNC), to replace retransmission mechanisms.     

Several RLNC variations and applications have been proposed~\cite{osel2023, 8554264,tasdemir2022fsw,pandi2017pace, wunderlich2017caterpillar,8633839}, but face major limitations of 
\begin{enumerate*}[i.)]
    \item being tailored to specific NC schemes and use cases, such as using the Sliding Window scheme for low-latency scenarios~\cite{tasdemir2022fsw,wunderlich2017caterpillar} or macro-symbols to reduce RLNC padding overhead~\cite{8633839};
    \item  limiting their approaches to end hosts (encoders and decoders) and rarely leverage the network by deploying recoders on network nodes, foregoing RLNC performance~\cite{tasdemir2021sparec,DsepVu,gabriel2018practical}; and 
    \item only simulating the network~\cite{tasdemir2021sparec,DsepVu} or using limited support~\cite{gabriel2018practical}, deploying recoders on servers rather than actual network switches or routers.
\end{enumerate*}
Till date, no studies have proposed RLNC for cloud-native 5G and beyond networks due to the challenge of integrating RLNC without disrupting existing architectures or deployments.

In this work, we employ \textit{programmable network coding} in cloud-native 5G and beyond networks enabling versatile RLNC operations to address these challenges.  Our approach offers the flexibility to use various RLNC techniques, encourages the adoption of RLNC protocols, and promotes the widespread use of RLNC protocols through \textit{seamless integration into already existing network deployments}.  
Specifically, we introduce Flexible Network Coding (FlexNC), which can combine different RLNC protocols flexibly. Given that packet loss—caused by factors such as lossy channels or network congestion—varies frequently, we can adapt FlexNC accordingly. For instance, during high packet loss, FlexNC directs traffic to an efficient RLNC protocol such as Sliding Window while under low packet loss conditions, FlexNC can switch to a more suitable RLNC protocol or even bypass RLNC altogether, directly forwarding traffic to the next hop to minimize latency.
We propose Recoder in the Network (RecNet) that leverages intermediate network nodes to join the coding process to increase the overall network programmability even further.  As RecNet employs in-network computing, it enables network devices such as routers or switches to recode encoded packets. 

\begin{figure}
\centering
  \includegraphics[width=\linewidth, clip]{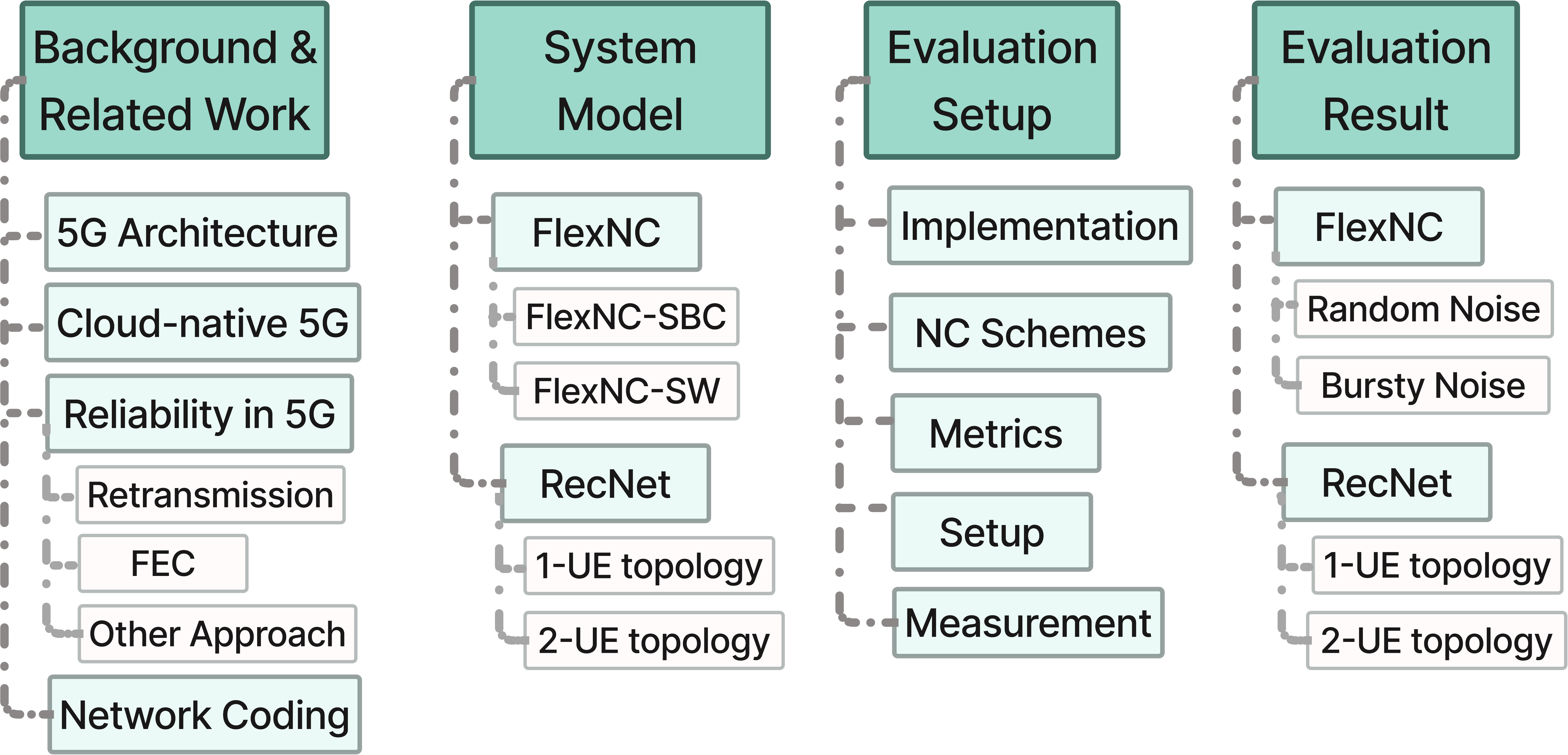}
  \caption{Structure of the paper}
  \vspace{-3mm}
  \label{fig:journal-structure}
\end{figure}

The overall structure of this paper is illustrated in Fig.~\ref{fig:journal-structure}, which provides the overall outline leading to the system description and performance evaluations.
We evaluated FlexNC and RecNet on a practical testbed, deploying the recoder on an actual network device enabled with in-network computing. 
To the best of our knowledge, we are the first to deploy and evaluate the recoder on an actual network device.  
We assess the performance of FlexNC and RecNet for OpenAirInterface (OAI)~\cite{oai}, a well-known deployment of a cloud-native 5G and beyond network and compare it against various benchmark schemes under different types of realistic traffic. 
While FlexNC significantly reduces the latency while still ensuring reliability, RecNet implemented as a Docker container on a Cisco High-performance Catalyst 9500 switch acting as a relay node significantly improves reliability while introducing insignificant latency.

\section{Background and Related Works} \label{sec:background}

In this section, we provide additional background for the overall system and provide an overview of related works. 
We describe the 5G Network Architecture used in Section~\ref{sec:background:5g-arch} and its cloud-based implementation in~\ref{sec:rel-work-5G}.  We detail retransmission and other approaches to generate reliable 5G RAN communications in Section~\ref{sec:background-retrans} before presenting network coding to provide reliability in Section~\ref{NC-background}.

\subsection{5G Network Architecture} \label{sec:background:5g-arch}
Two deployment modes are considered in 5G systems: Non-standalone (NSA) and Standalone (SA). By utilizing the 5G New Radio (NR) interface, both SA and NSA can provide functions and features that adhere to the guidelines established by the Third Generation Partnership Project (3GPP). The NSA architecture consists of a 5G RAN operating on a legacy 4G Long-Term Evolution (LTE) core. This architecture offers 4G and 5G base stations, with the 4G base station taking priority. On the other hand, 5G SA operates independently of existing 4G LTE infrastructure and the 5G NR is connected to the 5G Core Network (CN).  
5G SA is built from the ground up as a standalone network architecture optimized for 5G capabilities and accommodates new use cases and technologies (such as smart factories)~\cite{9163402}.  Therefore, we employ the 5G SA network architecture comprising a 5G CN and a 5G Radio Access Network (RAN).  The 5G RAN connects New Radio-User Equipment (NR-UE)\footnote{In this work, NR-UE and UE are used interchangeably.} to the 5G CN through the gNodeB (gNB). The gNB is the base station that communicates with the UE over the air interface and handles radio resource management, modulation, and beamforming functions. We employ the minimalist 5G CN deployment~\cite{mini-cn}, noting that our solutions apply to 5G CN in general. As illustrated in Fig. \ref{fig:sys-model-flexNC}, the minimalist 5G CN contains:

\begin{itemize}
    \item Network Repository Function (NRF): 
    enables 5G Network Function (NF) to register and deregister. It provides updates to NF and their services. 
    \item User Plane Function (UPF): connects to gNB via N3 interface on which user data is sent to UPF from RAN. UPF connects user data to external data networks (ext-DN) via the N6 interface. 
    \item Session Management Function (SMF): manages interaction with the decoupled data plane and handles session context with UPF through the N4 interface.
    \item Access and Mobility Management Function (AMF): manages UE registration, connection and mobility. AMF connects to gNB via the N2 interface to handle control-plane signalling. 
\end{itemize}

For this work, we implement a cloud-native 5G System consisting of one network to support control traffic between 5G components, another network for bridging user traffic between NR-UE and ext-DN, and wireless relay networks to provide coverage for remote UEs.

\subsection{Cloud-native 5G} \label{sec:rel-work-5G}
The cloud-native properties of 5G systems play a crucial role in enabling network operators to build and operate more agile, scalable, resilient, and cost-effective networks, typically deployed using containers. 
For example, 5G network automation in a cloud-native environment using Kubernetes to orchestrate and deploy containers was described in ~\cite{9110392} using OAI~\cite{OAI-5G}.  
Softwarized 5G core networks based on containers such as free5GC~\cite{free5G} and Open5GS~\cite{open5gs} were used in deployments such as 5G-Kube~\cite{scot2023} or in~\cite{9817268}, respectively. 
Similarly, the authors in~\cite{9345262} create a cloud-native version of the AMF. 

There have been works that cover the potential of cloud-native 5G RAN~\cite{nikaein2017towards}. In \cite{larrea2021nervion}, authors propose a scalable and flexible cloud-native RAN emulator, called Nervion, through abstractions and RAN element containerization. The authors in \cite{9269057} two vBBUs are aggregated into a single switched Ethernet Xhaul to assess the CloudRAN architecture. In \cite{10318161}, the authors analyze the micro-service architecture of RAN software components after introducing cloud-native principles and the state of RAN cloudification. The srsRAN~\cite{srsran} is an open-source project focused on creating a software-defined radio (SDR) implementation of the RAN components of mobile communication systems, particularly for 4G LTE and 5G NR (New Radio) standards. 

The paper in \cite{9481826} presents an Industry 4.0 application to show how the cloud-native principle in 5G can support heterogeneous vertical applications. In \cite{9482425}, introduces 5G-EPICENTRE which aims to utilize orchestration tools and container virtualization technologies to harness the performance gains that come with a fully containerized 5G Core network architecture. The authors in~\cite{9317860} propose an MEC-enabled network slicing using a cloud-native 5G microservices architecture. Whereas in \cite{8567377}, the authors use cloud-native principles and open-source commodity software to provide effective, multi-tenant access to mobile cloud resources of the Mobile Network Operator. They identify the challenges and the potential associated with switching the 5G cloud network's architecture from the old NFV design to the new cloud-native design.
The work in \cite{9877928} demonstrates a cloud network functions orchestrator-monitored end-to-end mobile network with channel emulation capabilities. It is a scalable, open-source solution that builds on popular cloud-native tools to simulate a whole 5G mobile network. \cite{9685874} employs an end-to-end real-world cloud-native deployment of a 5G network by exploiting the Kubernetes framework to analyze network utilization for either the RAN or the Core Network. 
The work in \cite{wissalVNF} offers a taxonomy of optimization strategies to solve the VNF placement issues in 5G. Particularly, they focus on the placement of Containers Network Functions in edge/fog computing. 
Significant research has been done on cloud-native 5G Systems but many of these do not focus on reliability in 5G. Our work aims to enhance 5G reliability by leveraging cloud-native properties. We first review current state-of-the-art reliability techniques in 5G and then present our alternative methods.

\subsection{General Reliability in 5G System} \label{sec:background-retrans}
Reliability is necessary for 5G networks to meet the diverse needs of users, businesses, and industries. 
One approach to achieve reliability is via retransmissions, as part of the current standardization efforts.  Additionally, we describe forward error correction and other approaches that exist to date before we focus on Network Coding in the section.

\subsubsection{Retransmissions (ARQ)}
The retransmission process between UEs and gNBs is crucial for ensuring reliable and error-free communication, especially in environments with poor signal conditions or high interference.
This process involves several layers of the 5G user protocol stack, including the Medium Access Control (MAC) layer and the Radio Link Control (RLC) layer, which play essential roles in managing and controlling the transmission of data between a UE and a gNB.  In downlink communications, packets are transmitted from gNBs to UEs with a gNB initiating data transmission by sending packets to the UE over the air interface.  First, the UE receives the packet (as air frame) at the physical layer (PHY) which is forwarded to the MAC layer, which manages the Hybrid Automatic Repeat Request (HARQ) process enabling error correction through retransmissions. The configuration parameters for HARQ, such as the number of transmission attempts and the timing for retransmissions, are set by the MAC layer. If the received packet is error-free and successfully decoded, the UE sends an acknowledgement (ACK) back to the gNB, otherwise a negative acknowledgement (NACK). 
The gNB's MAC layer awaits the ACK or NACK feedback. Upon receiving a NACK, the gNB's MAC layer initiates retransmission based on the HARQ configuration until an ACK is received or the maximum number of retransmission attempts was reached. 
After reaching the maximum number of retransmissions on the MAC layer, another retransmission can be triggered on the RLC layer.  The RLC  operates in three different modes, namely
i.) Transparent Mode (TM) without any retransmission or segmentation; 
ii.) Unacknowledged Mode (UM) where the RLC segments data received from the upper layers into Protocol Data Units (PDUs) or reassembles the received PDUs from the lower layer into complete data units (UM does not provide any acknowledgement mechanism for the received PDUs, thus, no retransmission is allowed); or
iii.) Acknowledged Mode (AM) where the RLC provides reliable data transmission using selective repeat Automatic Repeat reQuest (ARQ) mechanism. 
For reliable communication, the RLC layer of gNB and UE operate in AM mode. After the RLC transmitter at gNB performs segmentation/concatenation, it adds the RLC header and creates two identical copies. One copy of the data is sent to the MAC layer and another copy to its Retransmission buffer. The RLC packet (RLC PDU) in the retransmission buffer is retransmitted if the RLC (at gNB) receives a NACK or does not receive a response from the UE for a predetermined amount of time. Otherwise on receiving ACK, the RLC PDU in the retransmission buffer would be removed. The retransmission at the RLC layer can also continue until a maximum count is reached after which the connection between UE and gNB is dropped.
AM offers the highest level of reliability but introduces additional latency and overhead due to the acknowledgement mechanism and the possible increasing number of retransmissions.

\subsubsection{Forward Error Corrections (FEC)}
The Forward Error Correction (FEC) codes are extensively evaluated and presented as a method to provide reliability by enabling the detection and correction of errors that occur during data transmission. 
Additionally, the FEC codes are also used in conjunction with ARQ to further bolster reliability. 
Therefore, we will review works done with FEC codes and conjunction with ARQ since the repetition of packets and retransmission mechanism increase the delay significantly. 

Known end-to-end codes used for FEC are Reed-Solomon, Luby Transform (LT),  Raptor, and RaptorQ which is an enhanced version of Raptor code. 
Among all of them, RaptorQ has the best coding efficiency and authors in~\cite{FeiRaptorq} have presented RaptorQ code in the application layer of 5G for 5G Multicast Broadcast Multimedia Service (MBMS). 
They have done numerical analysis and the results show that video stream transmission has been improved with RaptorQ codes.
However, RaptorQ code is performing well if the generation size is high as shown in~\cite{DsepVu} and it does not permit mixing packets in relay nodes. However, high generations are not preferred for applications demanding low delay and it has shown in~\cite{shi2013whether} that mixing packets in the relay nodes improves performance. 
Therefore, considering the aforementioned aspects, we studied RLNC for 5G system since it is efficient for low delay and permitting the coding operations in the relay nodes.

\subsubsection{Other Approaches}
In addition to FEC-related solutions, the research on improving reliability in 5G systems encompasses a broad spectrum of work focusing on alternative methodologies. By exploring these diverse works, we aim to understand the current research done to address various aspects of reliability challenges in 5G and its use cases. 
In~\cite{8971961}, authors improve the reliability of a 5G network in an automated factory having mobile robot-mounted UEs by providing spatial diversity gains using numerous transmission/reception points (TRxPs) in an interference-limited environment. By carefully planning the number of TRxPs, their placement, and their cluster size, the authors can enhance availability, mean time to recovery, and mean time between failures for their industrial automation factory scenario. However, their investigations and results are based on simulations and not real testbed or emulation. 
The authors in~\cite{8314688} aim to provide end-to-end reliability for mission-critical traffic by introducing an NFV-based softwarized 5G architecture. They propose a mathematical methodology modelling the softwarized 5G RAN to manage mission-critical data from mobile high-priority users and develop a prototype hardware implementation to validate the important system design decisions.
In \cite{9482540}, the authors develop an algorithm, Make-Before-Break-Reliability, that includes application-network interaction and application server failure detection through an extension of the GPRS Tunnelling Protocol (GTP) header for ultra-reliable 6G network. In their algorithm, when an application server fails, the network recognizes the issue and redirects traffic to a backup route that connects to a redundant application server located in a different edge computing node.

\subsection{Network Coding Reliability for 5G Systems} \label{NC-background}
Network Coding (NC)\footnote{In this work, RLNC and NC are used interchangeably.} enables the intermediate nodes to combine packets known as the store-code-forward mechanism as an alternative to the store-and-forward mechanism. As a subversion of the NC, RLNC was developed in~\cite{Ho2006}. In basic RLNC, a coded packet is generated by the random linear combination of $N$ source packets. 
Afterwards, these coded packets are transmitted through the network to their destination where $N$ coded packets are required to decode $N$ source packets. RLNC does not necessarily need feedback from the destination to send packets. However, some signalling information should be appended for decoding. For instance, the encoding coefficients and the generation number are important information. 
Encoding coefficients are selected from a finite Galois field. If data has been divided into generations, then the generation number should also be known to group the received packets. 
RLNC does not only operate at the source, such as other end-to-end codes (e.g., Fountain codes) but allows recoding of the coded packets on relay nodes. 

The benefits of RLNC are boosted throughput~\cite{tasdemir2023computational}, improved reliability~\cite{lun2008coding}, security~\cite{LimaSec}, reduced network traffic in content delivery network~\cite{SandraComnets}, reduced need of energy~\cite{Vuenergy}, and distributed storage~\cite{JiangStorage}. 
Additionally, a centralized controller or knowledge of the network topology is not required for RLNC. Therefore, RLNC has been used for many different network scenarios, e.g.,  mesh networks~\cite{patrick}, ad hoc networks~\cite{nikolaos}, device-to-device communication~\cite{RLNCD2D}, and wireless IoT communication~\cite{rivera}. 
Variations of RLNC have been developed, e.g., i.) Fulcrum RLNC codes~\cite{tasdemir2022fsw, LucaniFulcrum} to address the heterogeneous devices with different computational capacities; 
ii.) sparce RLNC codes~\cite{tasdemir2021sparec, DsepVu} to address the complexity of the encoding and decoding process; or 
iii.) systematic block coding RLNC~\cite{Lucanisys} and sliding window RLNC~\cite{wunderlich2017caterpillar} to lower the delay of packets while being robust to the packet losses. In systematic block coding RLNC, a large chunk of the source packets are transmitted first, followed by redundantly coded packets. 
On the other hand, in sliding window RLNC, a small number of source packets are transmitted first, followed by redundantly coded packets to reduce delay.

\section{System Model} \label{sec:system-model}
We create an efficient and configurable 5G framework that implements network coding to enhance reliability and network performance. 
In this section, we present the overall system, including network coding (FlexNC) and recoding (RecNet) components.

\begin{table}[b!]
\caption{Definitions of abbreviations used in this work.} 
\centering
\begin{tabular}{|cc|}
\hline
\multicolumn{1}{|c|}{\textbf{Notation}}    & \textbf{Definition}\\ \hline
\multicolumn{1}{|c|}{$P$}    & Original packet\\ \hline
\multicolumn{1}{|c|}{$B$}    & Classifier's buffer\\ \hline
\multicolumn{1}{|c|}{$s_B$}  & Size of classifier's buffer\\ \hline
\multicolumn{1}{|c|}{$n_P$}  & Current number of original packet received by encoder  \\ \hline
\multicolumn{1}{|c|}{$t$}  & Type of traffic  \\ \hline
\multicolumn{1}{|c|}{$L_f$}  & List of ratio of traffic packets forwarded without NC  \\ \hline
\multicolumn{1}{|c|}{$\phi$}  & Randomly chosen slice of packets used for classifying $t$  \\ \hline
\multicolumn{1}{|c|}{$s_\phi$}  & Size of $\phi$  \\ \hline
\multicolumn{1}{|c|}{$T_s$}  & Sent timestamp of a packet  \\ \hline
\multicolumn{1}{|c|}{$T_r$}  & Receive timestamp of a packet  \\ \hline
\multicolumn{1}{|c|}{$I$}  & IPD between two packets \\ \hline
\multicolumn{1}{|c|}{$L_{T_s}$}  & List storing send timestamp of packets in $\phi$  \\ \hline
\multicolumn{1}{|c|}{$L_{\rho}$}  & List storing payload of packets in $\phi$  \\ \hline
\multicolumn{1}{|c|}{$L_I$}  & List storing IPD of packets in $\phi$  \\ \hline
\multicolumn{1}{|c|}{$\mu_I$}  & Average of IPDs of packets in $\phi$ \\ \hline
\multicolumn{1}{|c|}{$\mu_\rho$}  & Average of payloads of packets in $\phi$ \\ \hline
\multicolumn{1}{|c|}{$\sigma_\rho$}  & Standard deviation of payloads of packets in $\phi$ \\ \hline
\multicolumn{1}{|c|}{$G$}  & Generation size of SBC node \\ \hline
\multicolumn{1}{|c|}{$g$}  & Current generation number (id) \\ \hline
\multicolumn{1}{|c|}{$r$}  & No. of coded redundant packets in a generation of SBC node\\ \hline
\multicolumn{1}{|c|}{$q$}  & Galois Field size \\ \hline
\multicolumn{1}{|c|}{$E$}    & Encoder's buffer\\ \hline
\multicolumn{1}{|c|}{$n_E$}  & Current number of packet in encoder buffer\\ \hline 
\multicolumn{1}{|c|}{$w$}  & Window size of SW encoder\\ \hline
\multicolumn{1}{|c|}{$c$}  & No. of coded redundant packets generated after SW batch\\ \hline
\multicolumn{1}{|c|}{$b$}  & Batch of original packets of SW encoder\\ \hline
\multicolumn{1}{|c|}{$C$}  & Coded redundant packet generated by encoder\\ \hline
\multicolumn{1}{|c|}{$L_C$}  & List of redundant packets in an SBC generation or SW batch\\ \hline
\multicolumn{1}{|c|}{$R$}  & Packet received at decoder \\ \hline
\multicolumn{1}{|c|}{$D$}    & Decoder's buffer\\ \hline
\end{tabular}
\end{table}

\subsection{Programmable Network Coding: FlexNC} \label{sec:Flex-NC}

We propose a versatile RLNC algorithm in 5G Systems that supports the latency and reliability requirements of different types of traffic without modifying the existing RLNC protocols or the 5G infrastructure. 
This requires a classifier at the encoder to detect the traffic type of the incoming packets based on which the encoder customizes its operation. 
The challenge is to ensure a lightweight traffic classification that allows for real-time decision-making that is resource-optimized and suitable for large volumes of traffic. 
Another challenge is to synchronize the operations between the encoder and the decoder, for example to communicate the specific network coding protocol without increasing processing latency. 
The effectiveness and performance of the 5G system can be compromised by improper synchronization of operation of the encoder and the decoder leading to decoding errors, loss of redundancy, inability to regenerate original packets, ineffective error correction, increased latency, deteriorated QoS, unreliable communication, increased network traffic, and wasted resources.

\begin{figure*}[htbp]
\centering
  \includegraphics[width=\textwidth,height=6cm]{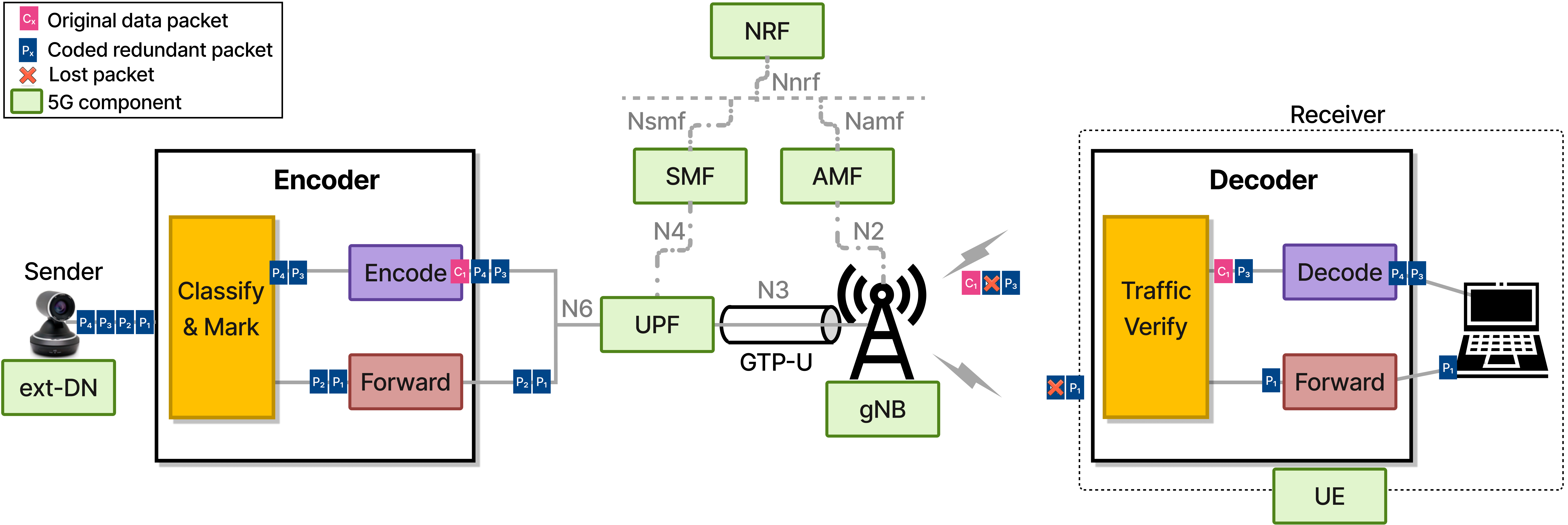}
  \caption{System model for the proposed algorithm: FlexNC}
  \vspace{-3mm}
  \label{fig:sys-model-flexNC}
\end{figure*}

In FlexNC, we combine multiple network coding nodes with the forwarder to implement a versatile RLNC algorithm in a 5G System.
Using multiple nodes provides a programmable and customizable environment and enables network operators to adjust the behaviour of individual nodes, deploy specific network coding schemes, or implement new features without disrupting the entire network. 
We illustrate FlexNC in Fig.~\ref{fig:sys-model-flexNC}. 
The main encoder block consists of the traffic classifier forwarding packets to either the encoder or the forwarder. Many recent network traffic classifiers use deep-learning models~\cite{8026581} and machine-learning algorithms~\cite{9321199}, which are prohibitive for FlexNC due to our real-time and resource constraints. 
Instead, we employ packet sampling for the traffic classification, similar to popular traffic classifiers such as sFlow and NetFlow. 
We specifically use random packet sampling involves selecting packets randomly from the total traffic to reduce the amount of data to be processed while still providing meaningful insights into network traffic patterns.  

Our approach ensures the fewest possible modifications for the current 5G standard protocol stack. 
As packet losses might occur even in the core~\cite{rischke20215g}, we place the encoder close to the ext-DN (sender) to create coded redundant packets ahead of the UPF. 
As indicated in Fig.~\ref{fig:sys-model-flexNC}, we install the decoder in the NR-UE to recover any lost original packets by decoding the coded redundant packets. 
FlexNC synchronizes the encoder and the decoder's operation by employing an "in-band" solution attaching the NC info to each packet, thus significantly reducing latency. 

Our proposed framework offers programmable network coding as an addition to the current 5GS architecture by integrating two network coding schemes with the forwarder in FlexNC: Sliding Window (SW) RLNC and Systematic Block Coding (SBC) RLNC. 
We choose these two schemes because they deliver low packet delay while guaranteeing reliability~\cite{tasdemir2021sparec}, \cite{9374634}. However, any other network coding scheme can also be integrated into FlexNC.

\subsubsection{FlexNC with SBC} 

In FlexNC we introduce multiple nodes with either network coding or forwarding capabilities. 
Data traffic is directed through either the forwarder or the network coding node as determined by the use case. We implement SBC-RLNC as the network coding protocol into FlexNC, referred to as FlexNC-SBC. 
The subsequent paragraphs discuss the encoding and decoding mechanism for FlexNC-SBC.

\paragraph{Encoding in FlexNC-SBC}

We outlined FlexNC's encoder workflow in Algorithm~\ref{alg:FlexNC-SBC-flow}, calling functions from Algorithms~\ref{alg:FlexNC-detect} and \ref{alg:FlexNC-SBC}. 
FlexNC-SBC has three encoding phases: Algorithm~\ref{alg:FlexNC-detect} covers the first two phases common to any network coding scheme, while Algorithm~\ref{alg:FlexNC-SBC} presents the mechanisms specific to FlexNC-SBC. 
We consider the downlink communication in 5G Systems. As illusrated in Fig.~\ref{fig:sys-model-flexNC}, packets from the sender at ext-DN traverse through the encoder, which detects and marks the type of traffic (Phases 1 and 2 of Algorithm \ref{alg:FlexNC-detect}).

\begin{algorithm}
\caption{Workflow of encoding in FlexNC}
\label{alg:FlexNC-SBC-flow}
	\renewcommand{\algorithmicrequire}{\textbf{Input:}}
	\renewcommand{\algorithmicensure}{\textbf{Output:}}
\begin{algorithmic}[1]
\Require $P$, $L_f$, $t$, $G$, $r$, $q$
\Ensure $P$, $L_C$  \Comment{Packet to be transmitted}
\State \textbf{Initialize:} $B \gets \emptyset$, $n_P \gets 0$
\Procedure{main()}{}
\While{True}
\State received $P$
\State $n_P \gets n_P$ + $1$
\State append($B$, $P$)
\If{$n_P \mod s_B = 0$} 
        \State $\phi \gets$ \textsc{rand\_slice}($B$) \label{call-rand-slice}  \Comment{See Phase 1, Alg 2}
        \State $t \gets$ \textsc{classify}($\phi$) \label{call-classify}  \Comment{See Phase 2, Alg 2}
    \EndIf
\If{$t = 1$}
    \If{$n_P > L_f[1]$} \Comment{Video traffic}
        \State $P, L_C \gets $ \textsc{encode}($P$, $t$, $G$, $r$, $q$)  \Comment{See Phase 3, Alg 3 (SBC) or Phase 3, Alg 5 (SW)}
    \Else
        \State $P \gets $ append($P.$payload, 41)
    \EndIf
\ElsIf{$t = 2$}
    \If{$n_P > L_f[2]$} \Comment{Haptic traffic}
         \State $P, L_C \gets $ \textsc{encode}($P$, $t$, $G$, $r$, $q$) 
    \Else
        \State $P \gets $ append($P.$payload, 42)
    \EndIf
\ElsIf{$t = 3$}
    \If{$n_P > L_f[3]$} \Comment{Audio traffic}
         \State $P, L_C \gets $ \textsc{encode}($P$, $t$, $G$, $r$, $q$)
    \Else
        \State $P \gets $ append($P.$payload, 43)
    \EndIf
\Else
   \State $P \gets $ append($P$, 0)
\EndIf
\State send $P$, $L_C$ to UPF and eventually to decoder
\EndWhile
\EndProcedure%

\end{algorithmic}
\end{algorithm}

The encoder listens for a packet, increments the packet count $n$ upon arrival, and adds the packet $P$ to the buffer $B$, which can store up to $s_B$ packets. 
Multiple packets are needed to classify the traffic type $t$, so a slice $\phi$) of $S_\phi$ packets is used. 
From every $s_B$ packets, a $\phi$ is randomly selected from $B$, necessitating a full buffer. When $n$ is divisible by $s_B$, the function to generate $\phi$ is called (Algorithm~\ref{alg:FlexNC-SBC-flow}, l. \ref{call-rand-slice}), commencing Phase 1 of Algorithm \ref{alg:FlexNC-detect}.

\begin{algorithm}
\caption{Traffic detection in FlexNC}
\label{alg:FlexNC-detect}
	\renewcommand{\algorithmicrequire}{\textbf{Input:}}
	\renewcommand{\algorithmicensure}{\textbf{Output:}}
\begin{algorithmic}[1]
\Require $P$, $s_B$, $s_\phi$, $B$
\Ensure $t$ 

\Algphase{Phase 1 - Select random slice from buffer}
\Function{rand\_slice}{$B$}
\State  $i \gets$ randint(0, $s_B$)
    \If{$i - s_\phi < 0 $}
    \State $\phi \gets B[i:i + s_\phi]$    
    \Else    
    \State $\phi \gets B[i - s_\phi:i]$
\EndIf
\State \textbf{return} $\phi$
\EndFunction

\Algphase{Phase 2 - Classify and mark traffic type}
\State \textbf{Initialize:} $L_{T_s}$, $L_{\rho}$, $L_I \gets \emptyset$, $t \gets 0$
\Function{classify}{$\phi$}
\For{$P$ in $\phi$}
    \State append($L_{T_s}$, $T_s(P)$)
    \State append($L_{\rho}$, $P$.payload)
    \EndFor
\For{($i=0;$ $i<s_\phi$-$1; i$++)}   
    \State $I \gets L_{T_s}[i$+$1]$ - $L_{T_s}[i]$ \label{calculate-IPD}
    \State append($L_I$, $I$)
    \EndFor
\State $\mu_I \gets$ mean($L_I$)
\State $\mu_\rho \gets$ mean($L_{\rho}$)
\State $\sigma_\rho \gets$ std\_dev($L_{\rho}$)
\If{($800 < \mu_\rho \leq 1500 $) $\&$  $\sigma_\rho<700$  $\&$ max($L_{\rho}$)$\geq 1300$ $\&$ min($L_{\rho}$)$>10$ $\&$ $\mu_I< 20$ } \label{if-video}
    \State $t \gets 1$ \Comment{Mark for video}   
\ElsIf{($L_{\rho}[$-$1]$-$100< \mu_\rho<L_{\rho}[$-$1]$+$100 $) $\&$  $\sigma_\rho<50$ $\&$ $L_{\rho}[$-$1]<350$}
    \State $t \gets 2$ \Comment{Mark for haptic} \label{if-haptic}
\ElsIf{($L_{\rho}[$-$1]$-$2< \mu_\rho<L_{\rho}[$-$1]$+$2 $) $\&$  $\sigma_\rho<0.4$ $\&$  max($L_{\rho}$)$< L_{\rho}[$-$1]$+$2$ $\&$ min($L_{\rho}$)$> L_{\rho}[$-$1]$-$2$}
    \State $t \gets 3$  \Comment{Mark for audio} \label{if-audio}
\Else    
    \State $t \gets 0$  \Comment{Mark for rest}
\EndIf
\State \textbf{return} $t$
\EndFunction

\end{algorithmic}
\end{algorithm}

In Phase 1, a random integer $i$ is selected between 0 and $s_B$. If $i \le s_\phi$, $\phi$ consists of packets from $B[i]$ to $B[i+S]$. If $i > s_\phi$, $\phi$ will consist of packets from $B[i-S]$ to $B[i]$. This $\phi$ is used to classify traffic (line \ref{call-classify}, Algorithm \ref{alg:FlexNC-SBC-flow}), triggering Phase 2 of Algorithm \ref{alg:FlexNC-detect}.

In Phase 2, the slice of packets is used to determine their traffic type $t$. This 4-byte unsigned integer is appended to each packet's payload. Initially, $t$ is set to 0 to indicate forwarding (no network coding). The sending time and the payload size of each packet in $\phi$ are stored in $L_{T_s}$ and $L_{\rho}$ lists respectively. The $L_{T_s}$ list is used to calculate the inter-packet delay (IPD) of $s_\phi-1$ packets (Algorithm~\ref{alg:FlexNC-detect}, l.~\ref{calculate-IPD}), and stored in the list $L_I$. The IPD mean ($\mu_I$), mean ($\mu_\rho$) and standard deviation ($\sigma_\rho$) of payload size are then computed to classify traffic. This work considers three common traffic types - video, haptic and audio. 

Video packets generally have higher, but variable, payloads. 
Due to the complex interplay of content characteristics, compression techniques, encoding parameters, and network conditions, video packets exhibit more variable payload sizes than audio and haptic. Video packets also have variable IPDs which are generally in the microseconds range. 
Algorithm~\ref{alg:FlexNC-detect}, l.~\ref{if-video} uses these statistics to detect video traffic, marking such packets with $t$ as 1. 
On the other hand, haptic packets typically have small, consistent payloads with a fixed 1~ms IPD and low variability, resulting in a lower standard deviation. Line \ref{if-haptic} of Algorithm \ref{alg:FlexNC-detect} detects haptic traffic, marking packet with $t$ as 2. Finally, we observed audio packets to have the least variability, with the lowest standard deviation in payload size. Line \ref{if-audio} of Algorithm \ref{alg:FlexNC-detect} detects audio traffic, marking packets with $t$ as 3. If none of these conditions are met, $t$ defaults to $0$. At the end of Phase 2, the obtained $t$ value determines the amount of network coding applied in Algorithm \ref{alg:FlexNC-SBC-flow}. 

The \textit{forwarding ratio} specifies the ratio of packets forwarded without network coding. Video, haptic and audio packets are marked with $t$ values of 41, 42 and 43 respectively, while other traffic types are marked with 0. The list $L_f$ contains forwarding ratios of different traffic types, configured by the network operator. A higher forwarding ratio means more packets are sent without network coding, reducing bandwidth consumption and latency but increasing packet loss. Operators must balance these trade-offs to meet QoS requirements efficiently.
After setting $L_f$ specific to $t$, the packet count is compared to $L_f$. For example, with an $L_f$ of 0.25 for video, the first 25\% of video packets are forwarded without network coding. 
If 1000 video packets are transmitted, the initial 250 video packets are forwarded and the remaining 750 video packets go through the SBC encoder (Algorithm~\ref{alg:FlexNC-SBC}). This constitutes Phase 3 of the FlexNC-SBC workflow.

\begin{algorithm}
\caption{FlexNC-SBC Encoder}
\label{alg:FlexNC-SBC}
	\renewcommand{\algorithmicrequire}{\textbf{Input:}}
	\renewcommand{\algorithmicensure}{\textbf{Output:}}
\begin{algorithmic}[1]
\Require $P$, $t$, $G$, $r$, $q$
\Ensure $P$, $L_C$

\Algphase{Phase 3 - Encode packet with SBC}
\State \textbf{Initialize:} $E, L_R \gets \emptyset$, $n_E, g \gets 0$
\Function{encode}{$P$, $t$, $G$, $r$, $q$}
\If {$n_E > G$}
    \State $n_E \gets 0$   \Comment{Reset packet counter}
    \State $E \gets \emptyset$  \Comment{Clear buffer $E$}
    \State $g \gets g + 1$  \Comment{New $g$}
\EndIf
\State append($E$, $P$) \Comment{Store packet in buffer $E$}
\State $n_E \gets n_E + 1$
\State $P \gets$ append($P.$payload, $g$) \Comment{Assign $g$ to $P$}
\State $P \gets$ append($P.$payload, $t$) \Comment{Assign $t$ to $P$}
\If {$n_E == G$}
\For{($i=r;$ $i<0; i--$)}   
    \State $R \gets \emptyset$
    \For{($j=g;$ $j<0; j--$)} 
       \State $R \gets R+$ randint(0,$q$)$E[j]$ \label{rand-sbc}    
    \EndFor
    \State append($L_C$, $C$) 
\EndFor
\EndIf
\State \textbf{return} $P$, $L_C$
\EndFunction
\end{algorithmic}
\end{algorithm}

In Phase 3, the SBC encoder~\cite{wunderlich2017caterpillar} operates in batches called generation, each consisting of original packets and coded redundant packets. 
Only the main idea of SBC encoding is presented here; a detailed description is available at~\cite{wunderlich2017caterpillar}.
The \textit{generation size}, $G$, determines the number of original packets per generation. 
When a packet arrives at the SBC encoder, its copy is stored in the encoder's buffer $E$. The packet belongs to a particular generation and is assigned a generation ID $g$.
The encoder also appends traffic type $t$ to the packet's payload. 
Then the packet is sent to the UPF. The buffer $E$ stores up to $G$ packets.
For example, if $G$ is 64, $E$ stores up to 64 packets and clears after every 64$^{th}$ packet.
Upon receiving the last original packet of a generation, the encoder generates $r$ coded redundant packets (e.g., $r = $ 32) by random linearly combining the original packets in $E$ using coefficients from a Galois field $GF(q)$ (Algorithm~\ref{alg:FlexNC-SBC}, l.~\ref{rand-sbc}). Each coded redundant packet is also appended with its $g$ and $t$. 
In the encoder, the payload size of the received packet is fixed to the largest traffic packet size, with smaller packets zero-padded.  
The last original packet of the generation is sent along with coded redundant packets to the UPF. After completing a generation of 64 packets, the encoder buffer's packet counter $n_E$ is reset and $E$ is cleared for the next generation, starting with a new $g$.

\paragraph{Decoding in FlexNC-SBC} \label{sec:decode-flex-SBC}
\begin{algorithm}
\caption{Workflow of decoding in FlexNC}
\label{alg:FlexNC-SBC-flow-decode}
	\renewcommand{\algorithmicrequire}{\textbf{Input:}}
	\renewcommand{\algorithmicensure}{\textbf{Output:}}
\begin{algorithmic}[1]
\Require $R$
\Ensure $P$
\Procedure{main()}{}
\While{True}
\State received $R$
\State $t \gets R.$payload$.t$  
\If{$t = 1$ OR $t = 2$ OR $t = 3$} \label{line:decode}
    \State \textsc{decode}($R$)
\ElsIf{$t = 41$ OR $t = 42$ OR $t = 43$ OR $t = 0$} \label{line:receive}
    \State $P \gets R$
\EndIf
\EndWhile
\EndProcedure%

\State \textbf{Initialize:} $D \gets \emptyset$
\Function{decode}{$R$}
\State append($D$, $R$)
\If{$D.full()$}
    \State $P \gets$  GE($D$) \Comment{Recover packets using GE()}
\EndIf
\EndFunction
\end{algorithmic}
\end{algorithm}

Once the encoder generates the coded redundant packets, it sends them along with the original packets to the UPF. The UPF encapsulates each packet with a GPRS Tunneling Protocol - User Plane (GTP-U) header, which includes routing like the Tunnel Endpoint Identifier (TEID) and the destination IP address. The UPF then forwards the packets to the appropriate destination via the RAN components. At the gNB, the GTP-U header is removed, and the payloads of the original and coded redundant packets are transmitted over the air interface to the UE. Some packets may be lost or corrupted during transmission.
The UE first receives the physical layer signals from the gNB over the air interface. It performs demodulation, channel estimation, and equalization processes to obtain the transmitted data symbols. Then, MAC layer processing identifies and extracts MAC layer headers. Higher-layer processing extracts the payload data for relevant applications. The decoder, at the application layer, processes packets before use (for example, to play video or audio).
As shown in Algorithm~\ref{alg:FlexNC-SBC-flow-decode}, when the decoder receives a packet ($R$), it first removes $t$ from the payload. The decoder uses $t$ to decide whether the packet is needed for decoding (line~\ref{line:decode}, Algorithm~\ref{alg:FlexNC-SBC-flow-decode}) or for direct use by the UE (line~\ref{line:receive}, Algorithm~\ref{alg:FlexNC-SBC-flow-decode}). Original and coded redundant packets are marked for decoding and stored in the decoder's buffer with the same generation ID ($g$). When all packets of a generation are received, the buffer is full, and the decoding process (e.g., Gaussian Elimination (GE)) begins. This recovers any lost packets of the current generation. The next packet with a new $g$ triggers the decoder to clear the buffer and start decoding for the new generation.

\subsubsection{FlexNC with SW}
To compare the performance of different network coding protocols in Flex-NC, we implement another scheme besides SBC. In FlexNC with SW, referred to as FlexNC-SW, the network coding protocol used is the SW-RLNC. In this subsection, we discuss the proposed FlexNC-SW algorithm.

\paragraph{Encoding in FexNC-SW}:
The workflow of FlexNC-SW is similar to FlexNC-SBC (Algorithm~\ref{alg:FlexNC-SBC-flow}) and includes three phases. Phases 1 and 2 of Algorithm~\ref{alg:FlexNC-detect}, detect and mark the traffic type of the received packet, are identical for FlexNC-SW. The difference lies in phase 3, involving the SW encoding process. In Algorithm~\ref{alg:FlexNC-SW-enc}, the SW encoder processes smaller batches (windows) of packets, storing copies in the buffer~\cite{wunderlich2017caterpillar} while sending originals to the UPF. The buffer size depends on the window size ($w$). Encoder parameters include the number of coded redundant packets ($c$) and the number of original packets ($b$) after which they are generated. For example, with $w$=16, $c$=8 and $b$=12, each of the 8 coded redundant packets is a random linear combination of the last 16 original packets. All 8 coded redundant packets are generated after every 12 original packets and sent to the UPF. As new packets arrives at the encoder, the window shifts, filling the buffer with the latest 16 original packets. This process detailed in Fig. 2 (b) of ~\cite{wunderlich2017caterpillar}.
When the first 12 original packets are received, the encoder's buffer, which holds 16 packets, can accommodate four more. Upon receiving the 16$^{th}$ original packet, the encoder  removes the oldest packet to store the new one, following a First-In, First-Out (FIFO) principle. After the next 12 original packets are received, the encoder generates 8 coded redundant packets using the 16 original packets in buffer: 12 original packets from the current batch and four original packets from the previous one. 
This continues till all the original packets marked to be encoded are encoded and transmitted to the UPF. 

\begin{algorithm}
\caption{FlexNC-SW Encoder}
\label{alg:FlexNC-SW-enc}
	\renewcommand{\algorithmicrequire}{\textbf{Input:}}
	\renewcommand{\algorithmicensure}{\textbf{Output:}}
\begin{algorithmic}[1]
\Require $P$, $t$, $b$, $c$, $w$
\Ensure $P$, $L_C$

\Algphase{Phase 3 - Encode packet with SW}
\State \textbf{Initialize:} $E, L_R \gets \emptyset$, $n_E \gets 0$
\Function{encode}{$P$, $t$, $b$, $c$, $w$}
\If {$n_E > b$}
   \State $n_E \gets 0$ \Comment{Reset packet counter}
\EndIf
\If {len($E$) $> w$}
    \State $pop($E$)$ \Comment{Pop oldest packet in $E$}
\EndIf
\State append($E$, $P$)
\State $n_E \gets n_E + 1$
\State $P \gets$ append($P$, $t$)
\If {$n_E == b$}
\For{($i=c;$ $i<0; i--$)}    
    \State $R \gets \emptyset$
    \For{($j=w;$ $j<0; j--$)} 
        \State $R \gets R+$randint(0,x)$E[j]$  
    \EndFor
    \State append($L_C$, $C$)
\EndFor
\EndIf
\State \textbf{return} $P$, $L_C$
\EndFunction

\end{algorithmic}
\end{algorithm}

\paragraph{Decoding in FexNC-SW}
Similar to Section~\ref{sec:decode-flex-SBC}, at the UPF, the received packets are tunnelled to the gNB, which transmits them as air frames to the UE. After demodulating, decoding, and processing the physical layer signals, the UE extracts the payload data through upper layer processing. The decoder uses the payload from the original and the coded redundant packets to recover any lost packets, as detailed in Algorithm~\ref{alg:FlexNC-SBC-flow-decode}. Unlike SBC decoder, the decoder in FlexNC-SW operates over a smaller window of packets.

\subsection{Programmable Network Coding: RecNet}

\begin{figure*}
\centering
  \includegraphics[width=0.9\textwidth,height=3.5cm]{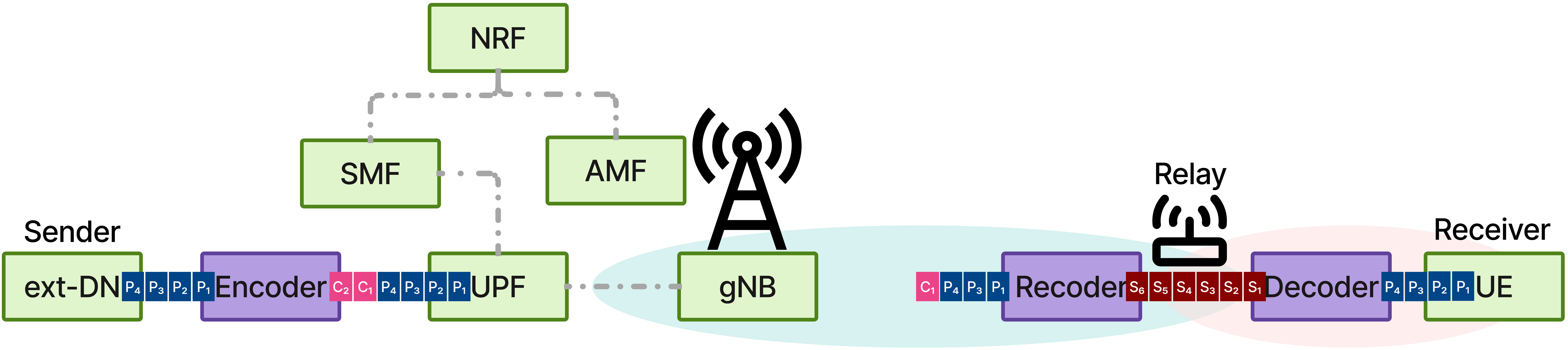}
  \caption{System model for the proposed algorithm: RecNet}
  \vspace{-3mm}
  \label{fig:sys-model-RecNet}
\end{figure*}

One of the primary objectives of 5G networks is to provide ubiquitous coverage, reaching even remote or sparsely populated areas. Remote UEs are crucial in industrial applications, such as mining, agriculture, or oil and gas, as well as for emergency services and public safety in disaster-prone areas. 
Addressing the connectivity needs of remote UEs is essential for creating an inclusive and robust 5G ecosystem that benefits users, industries, and communities. Our goal is to increase reliability for remote UEs that face communication challenges due to distance, obstacles, or terrain. To extend coverage, strategically positioned relay nodes are required to improve signal strength in areas with poor coverage.  
As shown in Fig.~\ref{fig:sys-model-RecNet}, the relay node receives signals from the gNB, amplifies them, and then retransmits them to the remote UE. Ensuring minimum packet loss in the wireless channel between the relay node and remote UE is challenging but essential for seamless communication. Significant packet loss between the relay and remote UE can degrade performance and user experience, even if the gNB-to-relay link is reliable. Without a reliable link to the remote UE, the advantages of deploying the relay node are diminished, leading to incomplete data delivery, degraded quality for specific UEs, disruptions in multicast and broadcast services, limited mobility support, and challenges for real-time applications with jitter and interruptions. Additionally, an unreliable link may force the relay node to use more energy and bandwidth, while UEs may need higher transmit power to maintain a connection, leading to faster battery drain.

One way to maintain reliability between the relay node and UE is retransmission, but this can significantly increase delay. Another option is to send additional redundant packets from the relay node, which can increase bandwidth consumption. We need a solution to improve reliability between the relay node and UE without adding high delay or bandwidth. 

The recoder random linearly combines the payload of the received packets and sends it to the decoder with minimum processing delay, without increasing the packet count. In wired networks, recoders can be easily deployed on routers and switches. However determining where to place the recoder in 5G networks is challenging, as it must not alter the existing architecture. Moreover, identifying beneficial scenarios for recoders in 5G networks is crucial; otherwise, they may add processing time without improving reliability.
Our proposed algorithm, RecNet, uses a smart relay node that recodes each received packet to include more information. This additional coded information helps the decoder recover packets lost between the relay node and UE.
To demonstrate the benefits, we will consider two 5G topologies with an integrated recoder. 

\subsubsection{One-UE topology} \label{Sec:1UE}

Similar to Section~\ref{sec:Flex-NC}, the one-UE topology considers the downlink communication from ext-DN sender to UE receiver. However, here the UE is far from the gNB, necessitating a relay node to enhance coverage. The relay node integrates a recoder to enhance reliability. The recoder is better suited for encoders using block coding protocols. Thus, RecNet uses only SBC, with the decoder at the UE.

As shown in Fig.~\ref{fig:sys-model-RecNet}, packets from ext-DN will first pass through the encoder. The encoding process follows Algorithm \ref{alg:FlexNC-SBC}, with a single encoder node random-linearly combining original packets to generate redundant coded packets. The user specifies $G$, $r$ and GF($q$) for the encoder. The encoder sends original and coded redundant packets to the UPF, where they are encapsulated with a GTP-U header and sent to the gNB.  
The gNB extracts the payload and sends it to the relay node via the air. 
The relay node can be a Wi-Fi 6 device because of its capacity to support high data rates, massive UEs connectivity and low latency device~\cite{batistatos2023wi}. In our proposed architecture, the relay node is connected to the gNB as a normal UE. 

\begin{figure}%[H]
\centering
  \includegraphics[width=0.93\linewidth, clip]{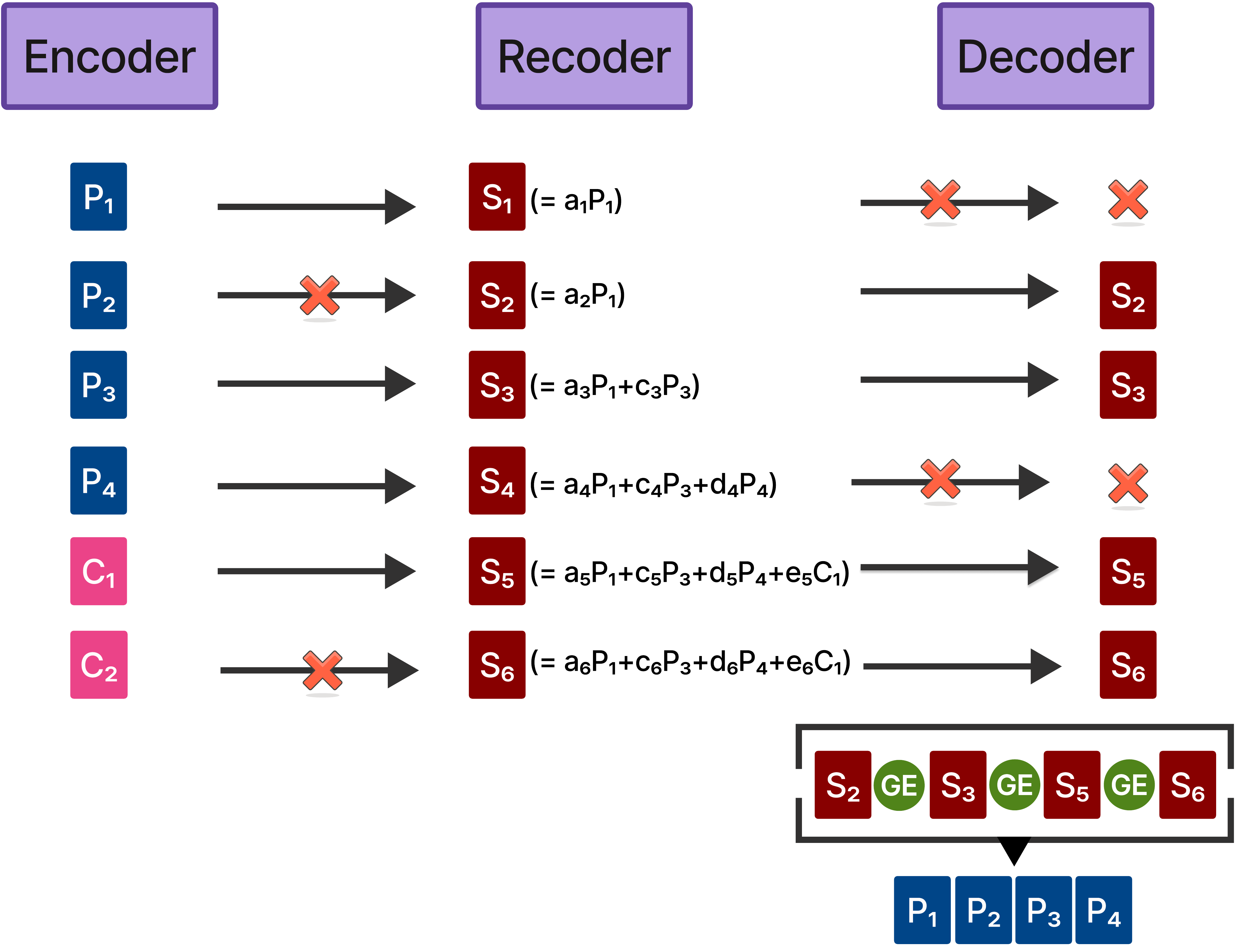}
  \caption{Basic example of Random Linear Network Coding in RecNet}
  \vspace{-3mm}
  \label{fig:recoder}
\end{figure}

At the relay, the recoder receives original and coded redundant packets, storing a copy of each in its buffer and forwarding their random linear combination. An example in Fig.~\ref{fig:recoder} shows the recoder receives the original packet $P_1$ and sends a coded combination $S_1$. If the next original packet $P_2$ is lost, the recoder forwards $S_2$, another combination based on $P_1$. Upon receiving $P_3$, it combines $P_1$ and $P_3$ to send $S_3$. This process continues until the last packet of the generation is received, then the buffer is cleared for the new generation. The recoded packets are sent to the UE, where the decoder stores them and uses GE to recover the original packet ($P_1$, $P_2$, $P_3$ and $P_4$) once all coded packets ($S_2$, $S_3$, $S_5$ and $S_6$) from the generation are received.

\subsubsection{Two-UE topology}
The next topology for RecNet is the two-UE topology, where communication occurs between two UEs through 5G links, involving both uplink and downlink. In this setup, one UE sends packets to a remote UE via a relay node, similar to Fig.~\ref{fig:sys-model-RecNet}. The receiver UE, relay node, recoder and decoder remain the same, but the sender is another UE within the same gNB coverage. The encoder is integrated at the sender UE and generates coded redundant packets using the SBC RLNC protocol (like in Section~\ref{Sec:1UE}).
The sender UE transmits these packets over the air to the gNB. The gNB forwards to the relay, where the recoder performs recoding as shown in Fig. \ref{fig:recoder}. The resulting coded packets are sent to the receiver UE, where decoding similar to Section~\ref{Sec:1UE} is done.

\section{Evaluation Setup}
In this section, we describe the overall system configuration and setup, including the determined performance metrics.
\subsection{Implementation}

A smooth transition to evaluation over a testbed or production network can be achieved by using RAN emulation~\cite{larrea2021nervion}. For the implementation of our proposed system model, we use the OpenAirInterface (OAI) software~\cite{OAI-5G} to deploy OAI RAN (OAI RF Simulator), 5G OAI UE, and 5G CN. 
We choose OAI to implement our cloud-native 5G and beyond network because it is an open-source project offering Standalone 3GPP compliant 5G stack implementation of CN, RAN, and UE, providing a complete platform for developing cellular systems. 
Other popular open-source projects only provide either the CN or the RAN. For example, open-source projects like srsRAN~\cite{srsran} provide a complete 5G RAN solution but exclude a UE application and CN, and UERANSIM~\cite{ueransim} offers 5G UE and RAN (gNodeB) simulator which does not include CN. Meanwhile, Open5GS~\cite{open5gs} and free5GC~\cite{free5G} are open-source projects that provide only 5G CN.
Each OAI 5G component is implemented using Docker containers and connected using three networks: two networks as Docker bridges to 
(i) manage inter-5G component control traffic and 
(ii) represent the external network connected to UPF from which user data is forwarded to the UE and a third network to (iii) represent the network connecting the relay node and UE. 

We employ the KODO library~\cite{kodo} to implement our solution with the NC Sliding Window and Systematic Block Codes protocols.
KODO is a C++ library which is designed to be utilized in real-world research on network coding protocols. We employ the KODO library because it offers personalized solutions, like generating distinct coding schemes. We use the Python wrapper on top of KODO to implement the NC protocols because of the user-friendly interface and ease of using Python. This method makes the C++-based KODO library more usable and accessible within the Python ecosystem by speeding up the development and prototyping process, improving cross-platform compatibility, and utilizing Python's extensive library ecosystem.
All the involved network coding nodes (encoder, recoder and decoder) must share the same values for the parameters, which include the payload size and the Galois Field (GF) size. 
The encoder's and the recoder's coded packets cannot be decoded by the decoder without this setting. The Sliding Window mode additionally requires the window size of the original packets used for producing coded redundant packets and the number of coded redundant packets sent after a specific number of original packets. Lost and corrupted packets can be recovered using the original packets and the coded redundant packets. 
To employ the System Block Code mode, the number of coded redundant packets and generation size are required. 

Since we target cloud-native 5G Systems, we deploy the network coding nodes on a Docker container. 
Moreover, we prefer Docker containers over Virtual Machines (VMs) and bare-metal due to their resource efficiency, portability, and consistency. Compared to VMs, containers use fewer resources because they share the host system's kernel and encapsulate applications with all dependencies, guaranteeing consistency across different environments. Because of their quick deployment and scaling capabilities as compared to bare-metal and VM, instances can be quickly spun up to handle increasing loads. Version-controlled Docker images make updates and rollbacks easier, while Docker's process-level isolation improves security and stability. 
One possible implementation of FlexNC is running a traffic classifier separately and having multiple encoders and decoders, each deployed on a dedicated container. However, this implementation will hinder scalability. 
FlexNC encoder and decoder are implemented in Python to reduce complexity in deployment, configuration, and monitoring. 
Taking the example of FlexNC-SBC, its encoder code comprises of a traffic classification function \textit{classifier} which as input takes a random slice of $S$ packets from a buffer of $s_B$ received packets. From the received random slice of $S$ packets, \textit{classifier} detects their traffic type. Depending on the traffic type, the forwarding ratio is fixed by the operator. The next incoming $s_B$ packets will be forwarded till the received packet count reaches the set forwarding ratio. Each of the forwarded packet's payload will be appended with a byte signifying the traffic type and that it does not experience network coding. As soon as the received packet count is above the forwarding ratio, then the next arriving $s_B$ packets will pass through the SBC encoding function \textit{encode}. The \textit{encode} function consists of an SBC encoder object which has input settings such as GF size, payload size, number of redundant packets and generation size as mentioned in the previous paragraph. Similarly, each of these packet's payload will be appended with an integer byte to represent its traffic type and that it is encoded. Meanwhile, the FlexNC decoder code checks the ending byte of each received packet's payload to determine whether it requires decoding or is ready to be directly processed by the application. If decoding is needed then the packets are passed to the decoder function \textit{decode}. The \textit{decode} function contains a SBC decoder object having the same parameters as that of the corresponding encoder object. The SBC decoder object will recover lost packets using for example Gaussian Elimination.
RecNet is also implemented using Python for encoder, recoder, and decoder. In the case of RecNet, all packets are encoded, recoded and decoded. 
For instance, the SBC encoder code consists of an encoding function \textit{encode}, which takes the received original packet as input. The SBC encoder object in \textit{encode} function will randomly linearly combine the packets when its buffer is full or a generation is reached. Then the the coded redundant packets are transmitted along with the original packet. At the recoder, \textit{recode} function buffers each incoming packet (both original and coded redundant packet) and random linearly combines it with the packets already in the buffer. This coded packet is sent to the decoder. The recoder's buffer is cleared only when a new generation of packets arrives. Finally, the decoder consisting of \textit{decode} function will buffer the received packet till it receives a packet from the current generation and then it starts decoding, via Gaussian Elimination.

\subsection{RLNC Configuration Details}
Both evaluated RLNC schemes can enhance reliability while consuming low packet delay~\cite{pandi2017pace,tasdemir2021sparec}. We employed the following parameters for the two RLNC schemes during our evaluation:
\begin{description}
    \item{Sliding Window (SW)}: redundant coded packets are spread between small batches of original packets. The SW employed encoder generates eight coded redundant packets ($c$) after every 16 original packets ($b$). The redundant coded packets are random linear combinations of the previous 16 original packets resulting in the coding window size ($w$) of 16.
    \item{Systematic Block Coding (SBC)}: here the encoder transmits redundant coded packets after a block of original packets~\cite{tasdemir2021sparec}. The same parameters must be configured in both the encoder and the decoder. Namely, the number of coded redundant packets ($c$) is 32 with generation size ($G$) equal to 64 for both the SBC encoder and the decoder.
\end{description}

RLNC linearly combines original packets across a Galois field GF($q$). RLNC can yield the optimal throughput for sufficiently large Galois fields while having insignificant linear dependence of the coding coefficient vectors~\cite{wunderlich2017caterpillar}. 
We consider a high field size of GF(2$^8$) for the encoder and the decoder of SW and SBC. The payload size is fixed for the network coding node in SW and SBC, which is generally the payload size of the largest packet of traffic. For our video, haptic and audio traffic we fix the payload size of the encoder and the decoder as 1328\,B,82\,B, and 480\,B, respectively.

\subsection{Performance Metrics}
We consider the one-way-delay (OWD) and packet loss as main performance metrics for the evaluation of the overall system performance targeting Ultra-Reliable Low Latency Communications (URLLC).

\subsubsection{One-Way-Delay}
The time a packet needs to go from one end of the network to the other is known as network delay. One-way-delay (OWD) accounts for potential asymmetry in network paths, where the delay in one direction may differ significantly from the delay in the opposite direction. OWD thus reflects the end-to-end delay experienced by data packets from the sender to the receiver, providing a direct measure of the time taken for data to traverse the network.
In contrast, the Round-Trip-Time (RTT) of the end-to-end delay may not accurately reflect the delay experienced by data travelling only in one direction because of differences in delays for each path direction due to pathway asymmetry~\cite{gerla2002tcp}. In applications where communication is primarily unidirectional (e.g., video streaming or content delivery networks), OWD provides a more relevant metric for assessing performance as compared to the RTT~\cite{10.1145/3466167}.  

To accurately measure the OWD, the sent and received times for each packet are required. We accomplish this by altering the tcpreplay~\cite{tcpreplayweb} traffic generator so that whenever a packet is transmitted, the sending timestamp of that packet is inserted into its payload. 
We calculate the OWD of a packet $P_i$, where $i \in \{1,2,3, \ldots, N_{r}\}$ with $N_{r}$ is the total number of packets received, by deducting the sending timestamp ($T_{s}$) from the receiving timestamp ($T_{r}$) as shown below: 

\begin{equation}
    OWD(P_{i}) = T_{r}(P_{i}) - T_{s}(P_{i})   \\ 
\end{equation}

\subsubsection{Packet Loss}
Packet loss refers to the failure of one or more data packets to reach their destination within a network. 
Packet loss can occur at the core~\cite{rischke20215g} and RAN (due to its wireless nature) in a 5G System. 
Packet loss can have significant impacts on network performance and the quality of service (QoS) experienced by users, particularly in real-time applications. Packet loss may also cause reduced accuracy, efficiency, or even system failure in some 5G applications~\cite{8972843}, which can have a major negative influence on system reliability.
Leveraging packet loss as a performance metric offers comprehensive visibility, real-time monitoring, and proactive management capabilities, empowering operators to optimize the performance, reliability, and QoS delivered by 5G systems.
We count the total number of packets received ($N_{r}$) at the UE in our measurements and compare these to the number of packets sent ($N_{s}$), provided by the corresponding pcap file for each test scenario (audio, video, and haptic).  We determine the packet loss subsequently as
\begin{equation}
    Loss = \frac{(N_{s} - N_{r})\times100}{N_{s}}.
\end{equation}

\subsection{Experimental Setup}

\begin{figure}[t]
 	\centering
 	\includegraphics[width=1.0\linewidth, clip]{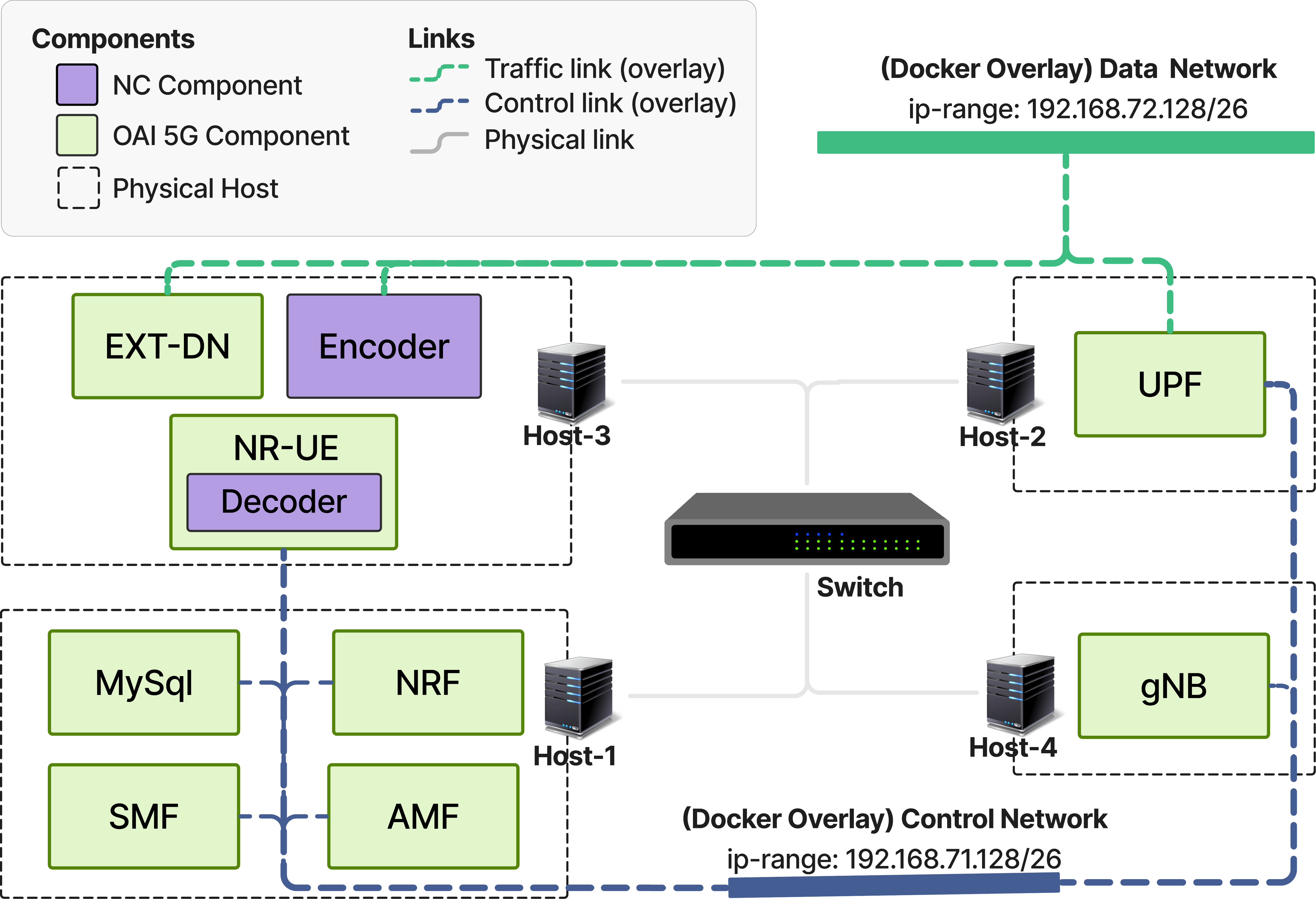}
 	\caption{Testbed architecture for the integration of RLNC into the OAI. All four host computers have an Intel Core i7-6700 (3.40GHz) CPU, 32GB RAM and  Ubuntu 18.04 OS. }
   \vspace{-2mm}
 	\label{fig:arch}
 \end{figure}

For realistic results, we configure a testbed that mimics environments used in production with multiple machines used to deploy different parts of the 5G system. Fig.~\ref{fig:arch} illustrates the architecture of our testbed which consists of four physical computers and a Gigabit switch. 
To replicate a real-world environment, we employed compact general-purpose computers in this testbed to closely simulate the actual setup, these multipurpose computers will provide the processing and the storage for the 5G components. Each computer features an Intel Core i7-6700 (3.40GHz) CPU, 32GB RAM, and two Network Interface Cards (NICs). One NIC is used for the management network to configure the computers while the second NIC is used by 5G components for control and data traffic.  Using multiple physical computers with the same hardware gives authentic results.

\subsubsection{Network}
To establish connectivity among the computers, we employed an Aruba 2930F switch equipped with Gigabit ports that deliver a throughput of up to 112.0 Mpps and a switching capacity of 176 Gbps\ as shown in Fig.~\ref{fig:arch}. 
This high-performance switching facilitates the transmission of both control and user traffic between 5G components.
Furthermore, we incorporated a Cisco Catalyst 9500 switch as an In-Network Computing (INC) device. The use of INC empowers network devices, such as switches and routers, to execute advanced tasks beyond their fundamental switching and routing functions. 
The Cisco Catalyst 9500 switch is equipped with a 4-core x86, 2.4-GHz CPU, 16 GBDDR4 memory, and 16GB internal storage. It provides extensive programmability and serviceability through its app hosting feature~\cite{app-hosting}, allowing the execution of various network applications as Docker containers.

\subsubsection{Software} \label{sec:eval-software}
All computers are operating on Ubuntu 18.04 (Kernel 4.15.0). Docker (version 24.0.1)~\cite{docker} is utilized on all machines to execute the OAI containers. Given that OAI involves multiple components that require specific configurations and a particular sequence of execution, Docker Compose (version 2.18.1) serves as an automation tool to streamline the deployment process.
The 5G and NC components, represented by containers, are interconnected through three distinct docker networks, the first two networks are docker overlay networks which were adopted by the OAI project and used as follows: i.) a control network that facilitates signalling traffic between the 5G components, ii.) a data network that manages user traffic to and from NR-UE, 
and iii.) a relay network established using the Docker MacVLAN driver~\cite{macvlan}. The latter network is dedicated to communication between the components operating on the computers and the recoder container running in the network on top of the Cisco switch. 

Given that random and burst packet losses may arise from factors, such as mobility or the intermittent nature of wireless channels, it is essential to employ a realistic simulator for noise to assess the impact of applying network coding during authentic packet loss scenarios in a 5G System. 
In OAI, the RF Simulator~\cite{oai-rf-sim} is used for providing the default noise model. However, we observed that while the noise model in the RF Simulator is stable in the AM mode, the noise model is still under development in the UM mode. This can cause unstable results when the RLC is configured in the UM mode. Therefore, we disable the default noise model in the RF Simulator and replace it with the Linux NetEm Server (tc-netem)~\cite{tc-netem} to simulate the necessary packet loss for various scenarios in the FlexNC algorithm. 

To comprehensively evaluate the performance of the FlexNC algorithm, we run the algorithm under different loss environments. We test the performance for i.) random uniform losses where packet loss is independent and ii.) bursty packet losses based on the Gilbert-Elliot model. 
The Gilbert-Elliott Model is popular amongst researchers for modelling bursty packet loss in the Internet~\cite{5755057}, wireless networks~\cite{7343635} and by the International Telecommunication Union (ITU) for capacity measurements on downlink access~\cite{ITU-TY1540}. We set tc-netem to have a total packet loss probability of around 10\,\% as considered in previous studies~\cite{tasdemir2023computational, tasdemir2021sparec, tasdemir2022fsw}.  
For the bursty loss model, we set the probability of starting a lossy state as 2\%, the probability of exiting the bad (lossy) state as 10\%, the packet loss probability in the bad state as 75\% and the packet loss probability in the good state as 0.1\% which resulted in around 12\% total packet loss probability.

In the RecNet algorithm, we employ the Cisco Catalyst 9500 switch to run the recoder. 
Due to switch limitations, we could not install the tc-netem package in this switch and, subsequently we use iptables to emulate a noise model between the source (ext-DN or sender UE) and recoder as well as between the recoder and decoder to evaluate the RecNet algorithm. 
We configured iptables to introduce a random packet loss at the recoder and destination node. The iptables does not consist of any noise model for bursty losses, thus, we only evaluate random loss for RecNet. 
For the experiments, we employed the tcpreplay (version 4.4.2)~\cite{tcpreplayweb}, a free and open-source tool, to reproduce the required traffic from captured PCAP files. tcpreplay, well-established in both research and industry, offers robust performance and provides numerous configurations. To support packet timestamping, we modified the tcpreplay source code, ensuring that it updates the packet timestamp to the current timestamp before sending the packet.

\subsubsection{Traffic data} \label{sec:eval-traffic}
5G networks are designed to cater to diverse applications with varying types of traffic and distinct characteristics. 
We captured and incorporated different actual content types to replicate authentic conditions and outcomes as PCAP files, namely traffic for 
\begin{enumerate*}[label=\roman*.)]
	\item audio characterized by the highest inter-packet delay (IPD) of 24 ms and a payload size of 480 Bytes, which falls between the packet sizes of video and haptic traffic (26,425 packets);
	\item video featuring variable IPDs ranging from microseconds to tens of milliseconds and the largest payload size of 1328\,Bytes (11,942 packets); and
	\item haptic exhibiting a low IPD of 1 ms and the smallest payload size at 82 Bytes (635,000 packets).	
\end{enumerate*}

\subsubsection{5G Deployment}
We used a Docker container to deploy each OAI~\cite{docker-yaml} component over different physical computers. 
We can achieve a more realistic outcome when the components are deployed on separate computers rather than sharing resources and operating on the same hardware. 
Every container is connected to either of the two or both overlay networks as discussed in Sec.~\ref{sec:eval-software}. 
In the FlexNC algorithm and one-UE topology of RecNet, the source of traffic is the external network that is deployed as the ext-DN container in OAI. To assess network coding in 5G, we deploy the encoder container to be in the same network as ext-DN and UPF by setting the IP address of the encoder container in the Docker Compose file. This helps to route the user traffic from ext-DN towards UE to traverse through the encoder container to UPF. 
We generate the encoder Docker image by creating an Ubuntu system Docker image in which the KODO-based Python libraries and the encoding program are installed. For the two-UE topology of RecNet, the source of traffic is a UE that we refer to as sender UE. In this scenario, we do not use a separate encoder container and install the encoder as an application in the sender UE container. 
 We achieve this by producing a Docker image with a KODO-based library included in the UE Dockerfile from OAI and running the encoding program in the sender UE.

In the proposed FlexNC topology, the destination UE is situated within the scope of gNB and does not require a relay node to expand coverage. 
Instead of creating a separate decoder container, we run the decoder as an application in the UE container. For the proposed RecNet topologies, the receiver (destination) UE is distant and out of coverage of the gNB, thus, requiring a relay node. Since in our system, the relay node functions as a normal UE to the gNB, we use the UE container from the OAI to represent the relay node. We connect the relay node container to the Cisco Catalyst 9500 switch. On this switch, we deploy the recoder Docker container. We configure all the packets forwarded by the relay node container to travel through the recoder container. To represent the destination node in RecNet topologies, we create a decoder Docker container that receives and decodes packets from the recoder container. 
With our suggested design, we enable flexibility in terms of available deployment possibilities for the network coding nodes. For example, the network coding node can be used as a Docker container if the UE is a laptop or PC equipped with CPU processors or an application if the UE is a smartphone. For accurate OWD measurement in FlexNC, we deploy the traffic source (ext-DN) container, encoder container, and destination receiver (UE) container on the same physical host as illustrated in the testbed in Fig.~\ref{fig:arch}. We use the same testbed for RecNet topologies wherein the sender UE, the relay node (second UE from OAI) and the decoder containers are located in the same host as ext-DN (i.e., host 3 in Fig.~\ref{fig:arch}). This is done to avoid having different timestamps in the containers.

\subsubsection{Measurement Process}
We evaluate the effectiveness of integrating network coding in a 5G System where retransmission is disabled. To prevent retransmission at the RLC and MAC layers, we configure the gNB to operate in UM mode and set the number of HARQ rounds to 1. 
We discuss the measurement scenarios for FlexNC and RecNet in the following paragraphs.

\paragraph{FlexNC}
The data flow in the user plane of the 5G System serves as the foundation for the measuring procedure. Each of the main measurement scenarios for the FlexNC algorithm is run 50 times for all pcap traffic (audio, video, and haptic), using a 95\% confidence interval based on the normal distribution. The variability in our results primarily stemmed from the packet loss simulation. During each test, no other processes were running except for the operating system, allowing our system to utilize all available resources. However, it's worth noting that we used Docker to run the 5G components, and this virtualization layer could potentially introduce some variability. We did not apply any offset to the traffic source, and we used the same traffic (pcap files) for each replication test to ensure consistency across runs. Every replication test consists of the following measurement scenarios: 
\begin{itemize}
    \item No NC: in this scenario, tcpreplay runs pcap files in the ext-DN container for generating traffic mentioned in Section \ref{sec:eval-traffic}. The packets destined for UE only pass through the 5G user plane, including UPF and gNB containers, to reach the UE container. There is no network coding node involved in this case.
    \item Pure NC: this measurement scenario consists of Pure SBC and Pure SW. Packets generated from ext-DN first pass through the encoder container where redundant coded packets are produced and sent along with the original packets. The resulting packets go through the UPF and gNB containers until they arrive at the UE container where decoding occurs. The decoder tries to recover any lost packet by using the received packets.
    \item FlexNC: this measurement scenario consists of FlexNC-SBC and FlexNC-SW. In this scenario, a portion of packets is directly forwarded to UPF without experiencing any encoding. The remaining packets are encoded. Similarly, UE receives the packets that were forwarded and decodes the packets that were encoded.
\end{itemize}

\paragraph{RecNet}
We evaluate the RecNet algorithm over two topologies. Each measurement scenario for both topologies for their respective traffic is run 10 times because the confidence intervals that were obtained were noticeably small. Because of the lower number of runs, we compute the 95\% confidence interval based on the student-t distribution. The main measurement scenarios for the RecNet algorithm are:
\begin{itemize}
    \item No NC: similar to the FlexNC measurement's No NC scenario, no network coding or retransmission is used. The packets of different traffic types originate from the ext-DN container and traverse the UPF, gNB, and relay containers to reach the remote UE container.
    \item Pure SBC: all packets sent from ext-DN are encoded at the encoder container. After gNB, the resulting original and coded redundant packets arrive at the relay node container. The relay container forwards the packets to the decoder which recovers the lost packets.
    \item RecNet: the relay node is the recoder container and recodes the packets received from the encoder. The recoded packets are then sent to the decoder where recovery of the lost packet takes place.
\end{itemize}

\section{Evaluation Results}
In this section, we present the results for FlexNC and RecNet algorithms and perform a detailed analysis. 

\subsection{FlexNC} \label{sec:eval-flex-nc}
We initially evaluate the performance of FlexNC under different noise models of 5G Systems. 
Additionally, these results will aid the analysis of which network coding scheme is more suitable depending on different noise conditions.

\subsubsection{Random Noise} 
In the following, we provide an evaluation of FlexNC in terms of packet loss and latency/OWD.

\begin{figure*}[t]
    \begin{center}
        \subfigure[]{
        \includegraphics[width=5.5cm]{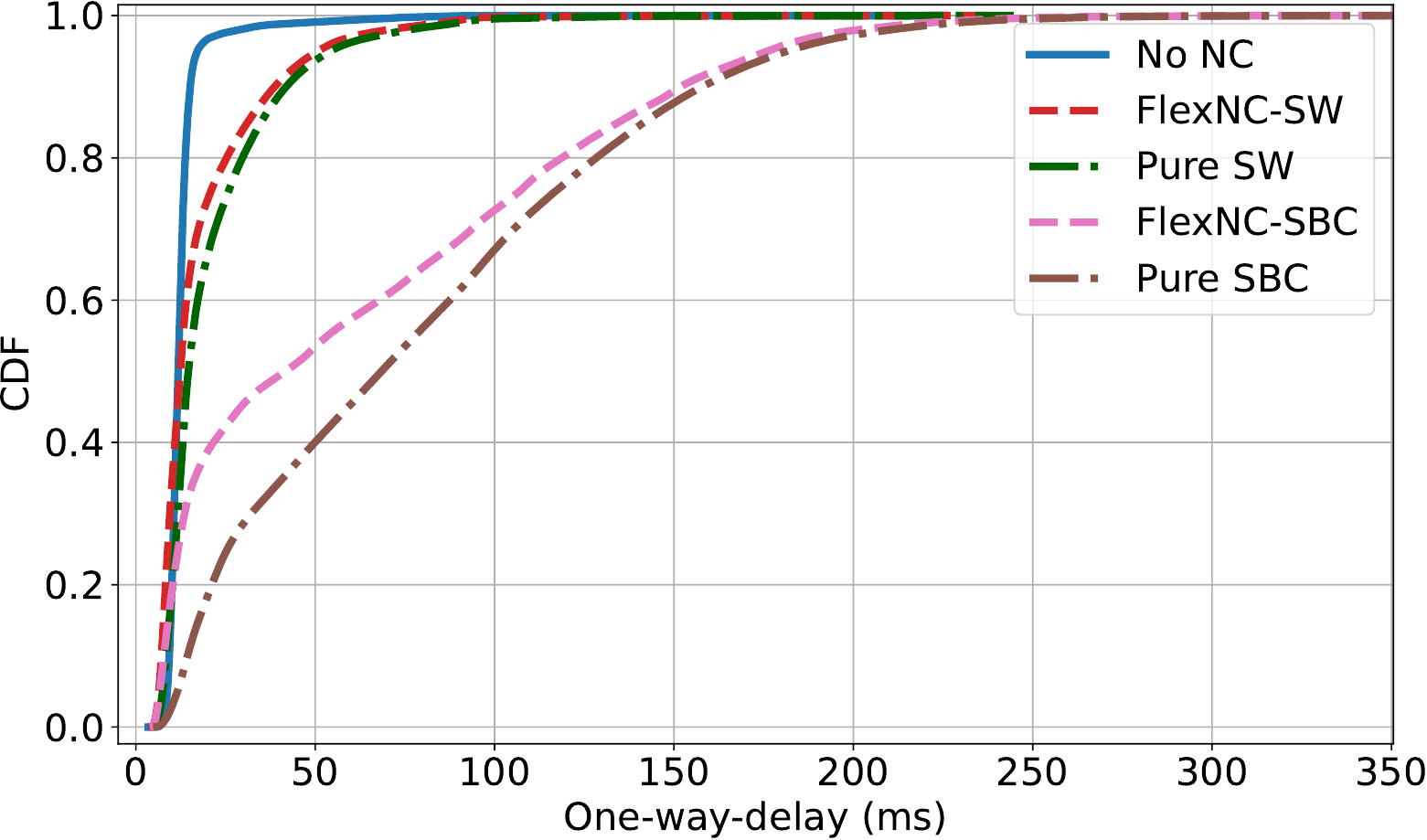} \hfill
        }
        \subfigure[]{
        \includegraphics[width=5.5cm]{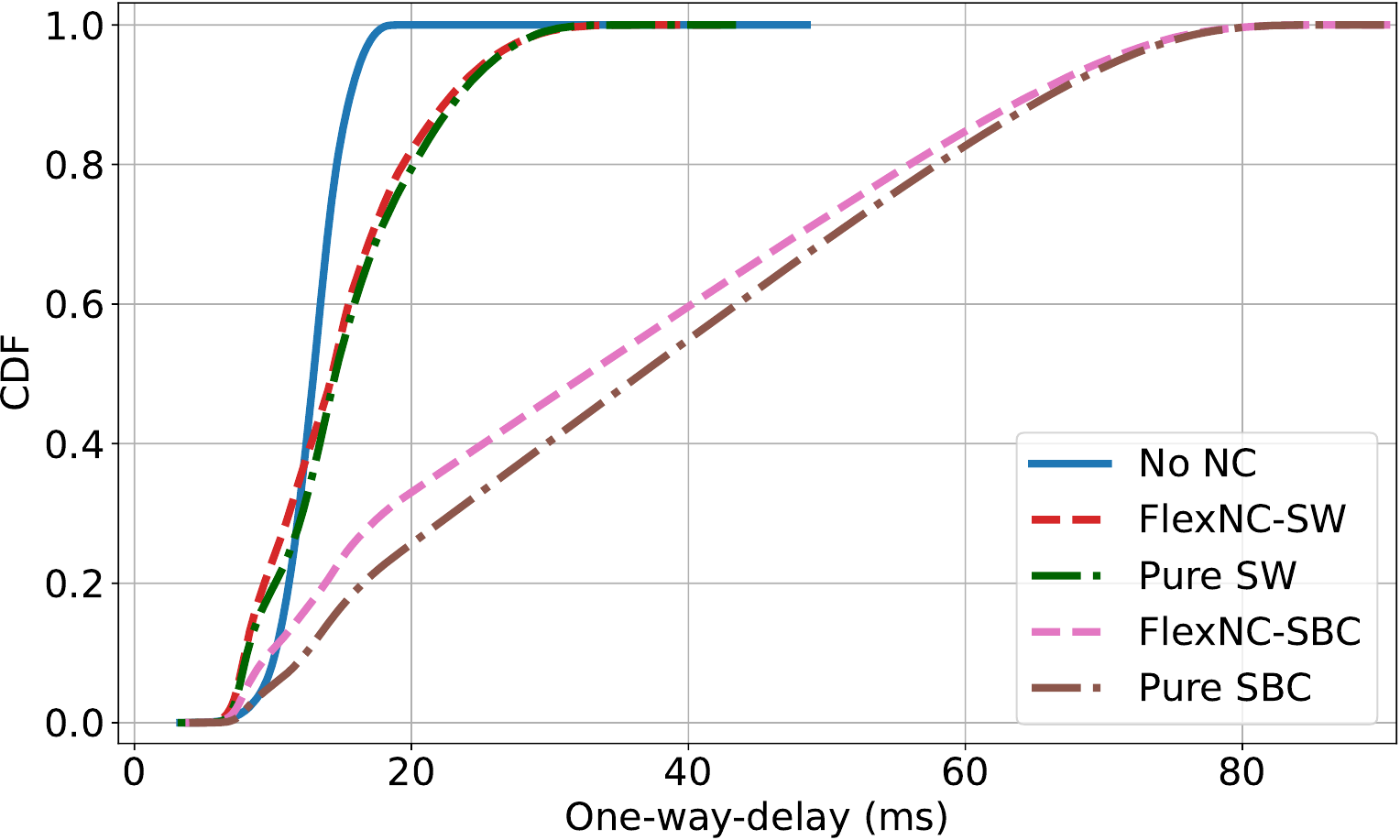} \hfill
        }
        \subfigure[]{
        \includegraphics[width=5.5cm]{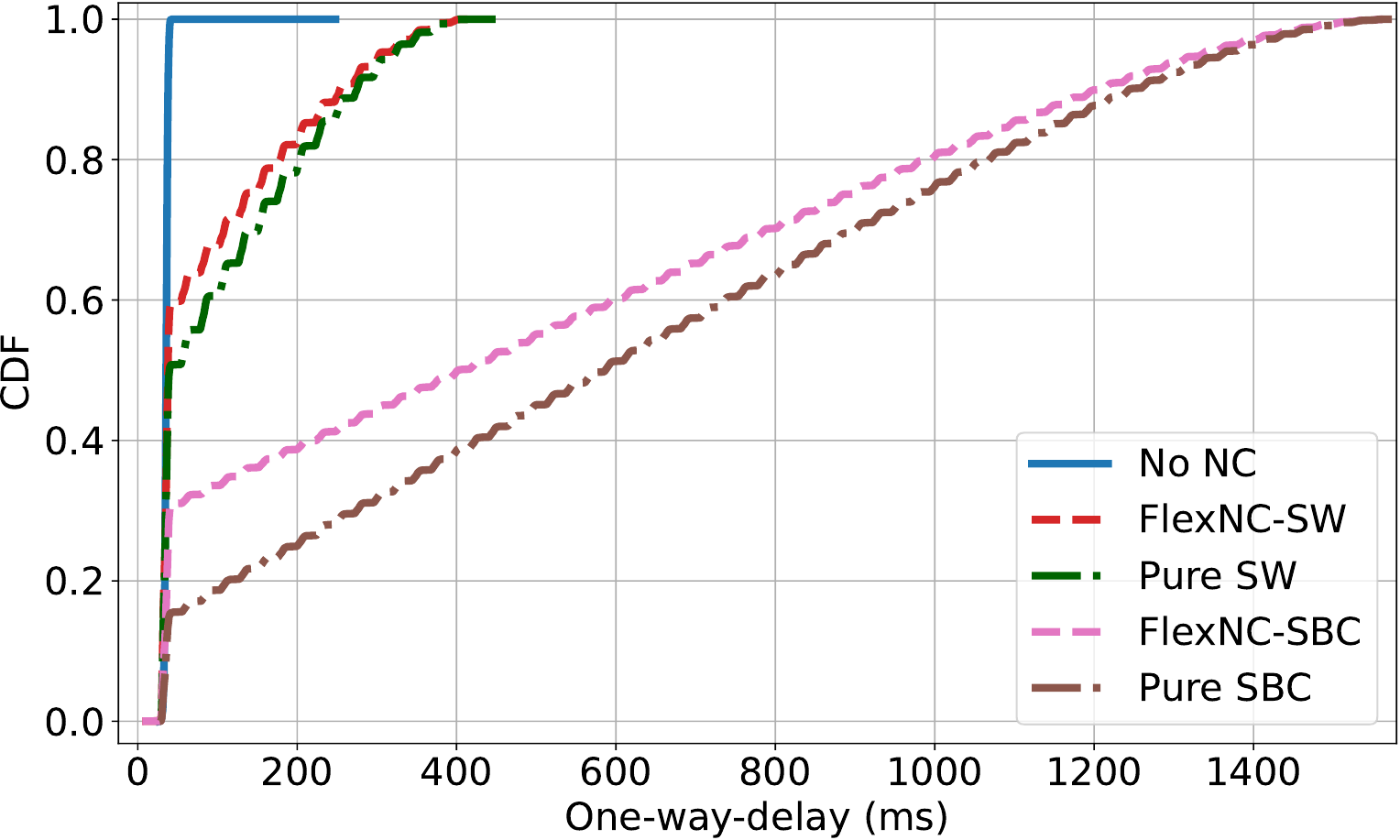} \hfill
        }
    \end{center}
    \caption{Combined overview of OWD measurements in downlink under random noise model for all received (a) video, (b) haptic, and (c) audio packets across all test runs.
    } 
    \label{fig:owd-flexnc-rand-TR}
\end{figure*}

\begin{figure*}[t]
    \begin{center}
        \subfigure[]{
        \includegraphics[width=5.5cm]{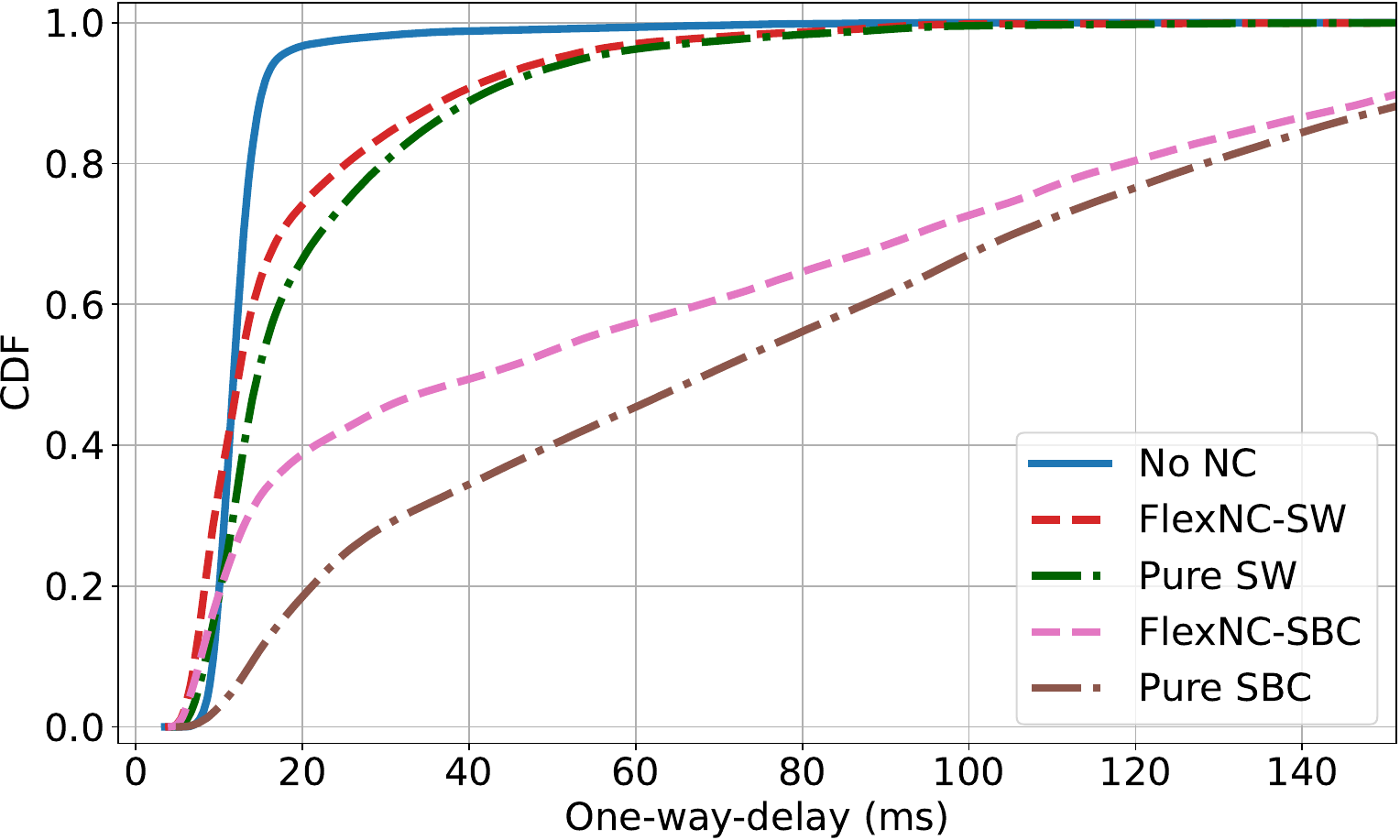} \hfill
        }
        \subfigure[]{
        \includegraphics[width=5.5cm]{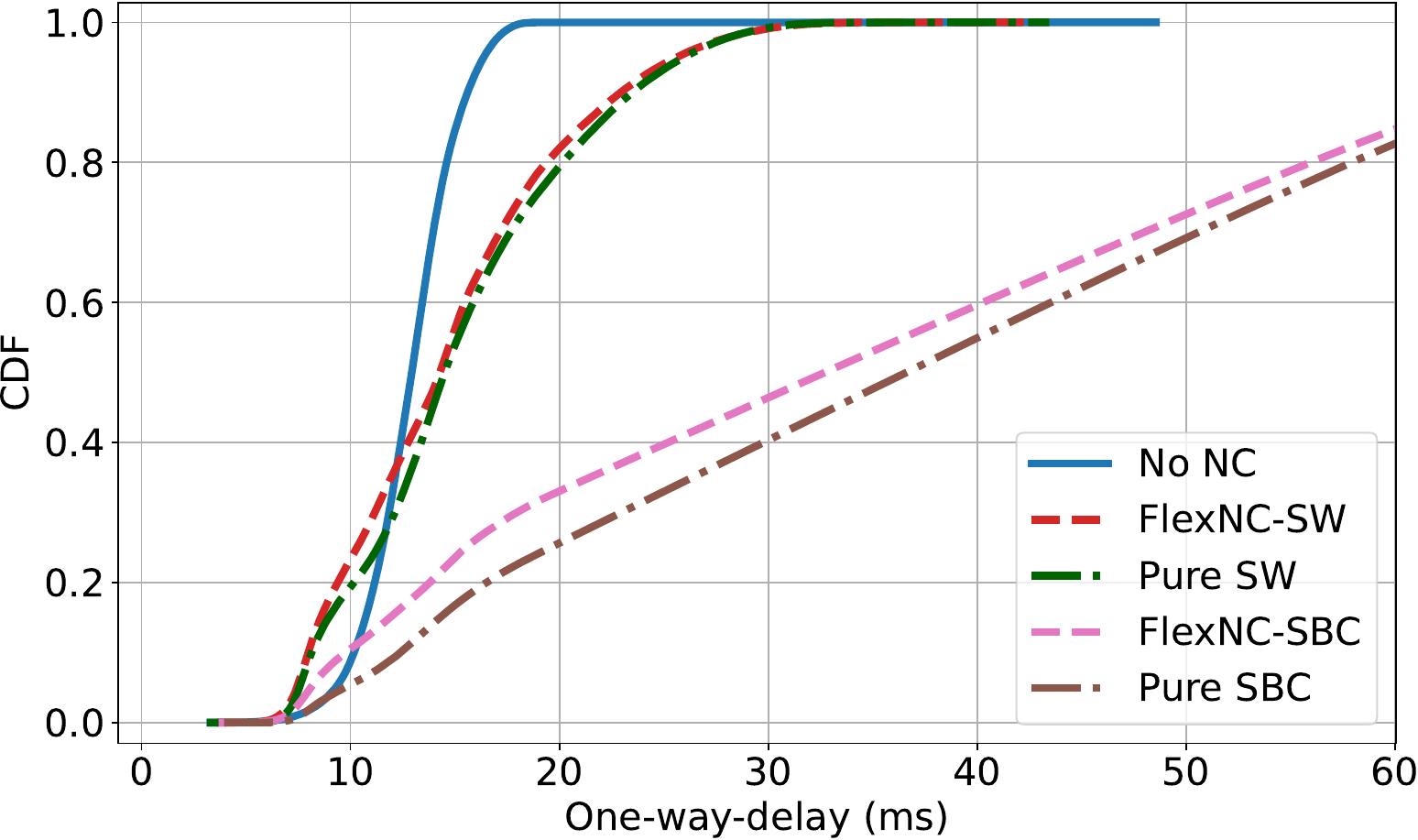} \hfill
        }
        \subfigure[]{
        \includegraphics[width=5.5cm]{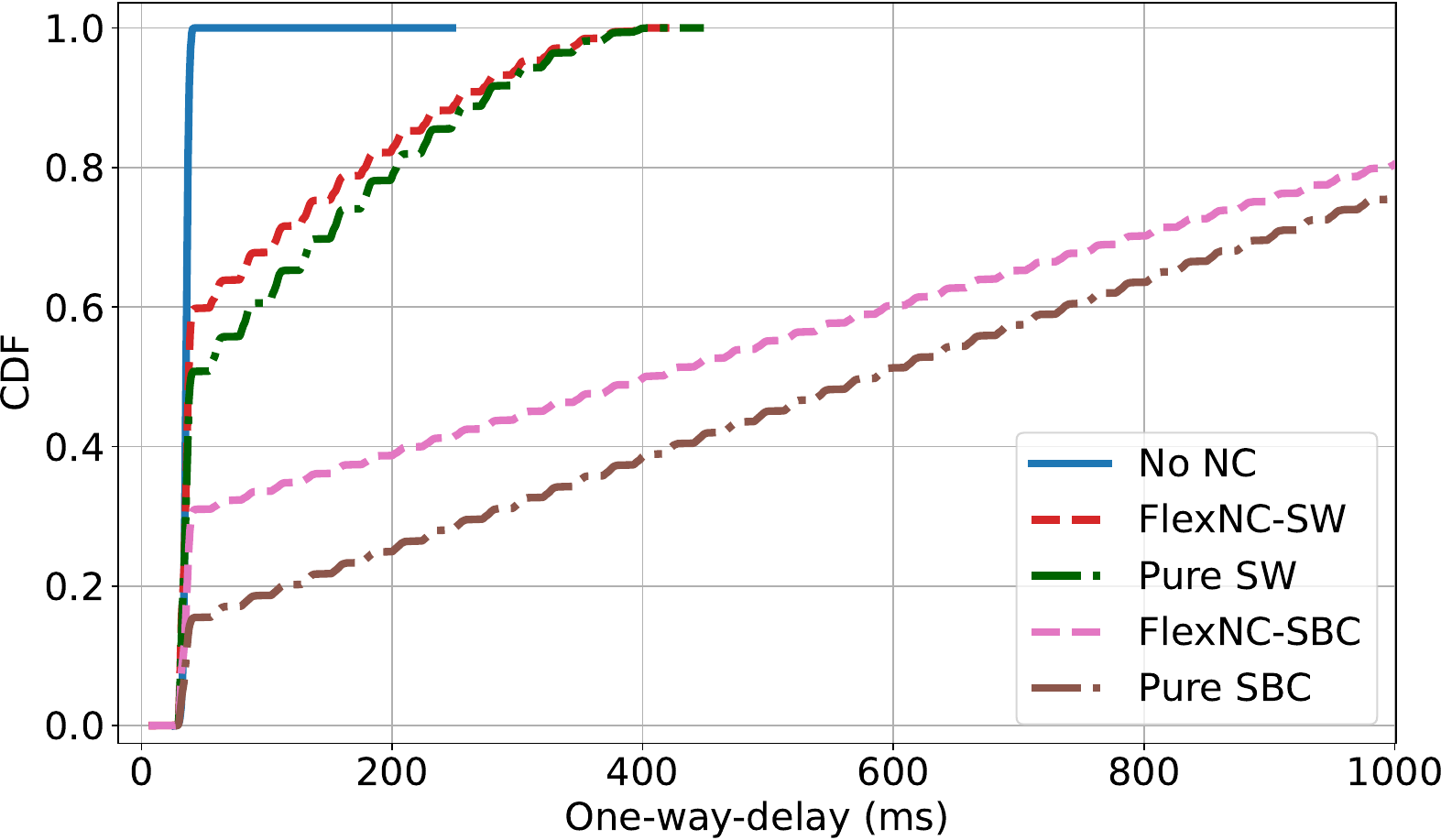} \hfill
        }
    \end{center}
    \caption{Combined overview of OWD measurements in downlink under random noise model for all received (a) video, (b) haptic, and (c) audio packets across all test runs, highlighting the region of interest from Fig.~\ref{fig:owd-flexnc-rand-TR}.}
    \label{fig:owd-flexnc-rand}
\end{figure*}

\paragraph{Latency / OWD} \label{sec:eval-rand-flexnc-owd}: 
We analyze how much additional delay is added when network coding is applied to the entire traffic and compare it with FlexNC. Fig.~\ref{fig:owd-flexnc-rand-TR}(a), \ref{fig:owd-flexnc-rand-TR}(b) and \ref{fig:owd-flexnc-rand-TR}(c) show the Cumulative Distribution Function (CDF) plots representing the OWD of all packets of video, haptic, and audio, respectively, across all 50 runs. Therefore, the plots do not include any CI.
The scenario without NC has the lowest OWD as there is no retransmission or redundancy applied, so the original packet is only sent once and immediately. 
We can further observe from Fig. \ref{fig:owd-flexnc-rand-TR} that the improvement in the OWD for FlexNC-SBC from Pure SBC is more prominent as compared to FlexNC-SW and Pure SW.
This is because the SW decoding time is lower than the decoding time in the SBC scheme.
While the redundant coded packet for SBC arrives at the tail of the generation, the redundant coded packets in SW are interleaved after every small batch of original packets. 
Subsequently, the waiting time at the decoder is higher for SBC than for SW. 
We additionally observe in Fig. \ref{fig:owd-flexnc-rand-TR} that the reduction in OWD with the application of FlexNC for haptic traffic is lowest compared to audio and video traffic because the ratio of the packet forwarded (without experiencing network coding) for haptic is the lowest at 10\%.  The overall impact of FlexNC is the lowest for haptic applications. On the other hand, we can see a noticeable decrease in OWD for video and audio, specifically for FlexNC-SBC compared to Pure SBC. 

For more detailed analysis, we observed in Fig. \ref{fig:owd-flexnc-rand-TR}(a) that 60\% of the received video packets for FlexNC-SBC have OWD value below or equal to 67.9\,ms while for Pure SBC 60\% of the received video packets have OWD value below or equal to 87.5\,ms. Thus, with FlexNC-SBC, 60\% of the received video packets can lower their OWD by 11.5\,ms. On the other hand, for FlexNC-SW, 60\% of the received video packets have OWD value below or equal to 14\,ms and for Pure SW the OWD are below or equal to 17.2\,ms. With FlexNC-SW, 60\% of the received video packets can lower their OWD by 3.2\,ms. When no network coding is applied then 60\% of the received video packets have OWD value below or equal to 12.3\,ms.
From Fig. \ref{fig:owd-flexnc-rand-TR}(a) we observe that the queue for the No NC scenario is built up as a result of the low IPD of video traffic, which causes the high OWD with almost all received packets obtained within 150ms. In the meantime, nearly all of the video packets can be received by Pure SW and FlexNC-SW in 200\,ms. Whereas Pure SBC and FlexNC-SBC can receive nearly all video packets in 300\,ms. 

Then we observed in Fig. \ref{fig:owd-flexnc-rand-TR}(b) that the OWD value of 60\% of the received haptic packets for FlexNC-SBC is below or equal to 40.3\,ms while for Pure SBC, OWD value is found to be below or equal to 43.5\,ms for 60\% of the received haptic packets. Therefore, 60\% of the received haptic packets can lower their OWD by 3.2\,ms by using FlexNC-SBC instead of Pure SBC. For FlexNC-SW, 60\% of the received haptic packets have OWD values below or equal to 15.5\,ms and for Pure SW, OWD is below or equal to 15.9\,ms. With FlexNC-SW, 60\% of the received haptic packets can lower their OWD by 0.4\,ms. Without network coding, 60\% of the received haptic packets have an OWD value below or equal to 13.4\,ms. 
Moreover, Fig. \ref{fig:owd-flexnc-rand-TR}(b) shows that for No NC scenario, almost all received haptic packets are obtained within 20ms. In the meantime, nearly all of the haptic packets can be recovered by Pure SW and FlexNC-SW in 35\,ms. Whereas Pure SBC and FlexNC-SBC can recover nearly all haptic packets in 85\,ms. 
We noticed a few outliers in haptic traffic which were excluded from our evaluation.

We observe in Fig. \ref{fig:owd-flexnc-rand-TR}(c) that the OWD value of 60\% of the received audio packets for FlexNC-SBC is below or equal to 593.2\,ms while the OWD value of 60\% of the received audio packets is below or equal to 736.1\,ms for Pure SBC. Using FlexNC-SBC in place of Pure SBC results in OWD value of 60\% of the received audio packets being reduced by 142.9\,ms. Whereas for FlexNC-SW, 60\% of the received audio packets' OWD values are below or equal to 54.7\,ms and for Pure SW the OWD values are below or equal to 86\,ms. Applying FlexNC-SW rather than Pure SW results in 60\% of the received audio packets' OWD being reduced by 31.3\,ms. Finally, we observed that 60\% of the received haptic packets without network coding have OWD value below or equal to 36.1\,ms.

As observed from Fig. \ref{fig:owd-flexnc-rand-TR}(c), the integration of network coding results in a higher OWD for audio traffic than the No NC scenario. In the case of No NC, most audio packets are received within 45\,ms. This is due to the high IPD (24\,ms) of audio packets which can be particularly detrimental to the SBC scheme. Our algorithms using SBC have $G$=64, so the coded redundant packets are generated and sent after 64 original packets. Thus, the SBC decoder has to wait for at least 64$\times$24=1536\,ms to receive the 64 original packets and coded redundant packets before it can start decoding. Meanwhile, for the SW scheme, the coded redundant packets are spread over a smaller batch of original packets ($b$). In our testbed, since $b$=16, the redundant coded packets are sent out after a batch of 16 packets, causing a waiting time (for audio packets) of 16$\times$24=384\,ms. Due to the high recovery time, Pure SBC and FlexNC-SBC need around 1550\,ms to receive most of the audio packet, while Pure SW and FlexNC-SW need around 400\,ms.

\begin{figure}[t]
    \centering
    \includegraphics[width=0.95\columnwidth]{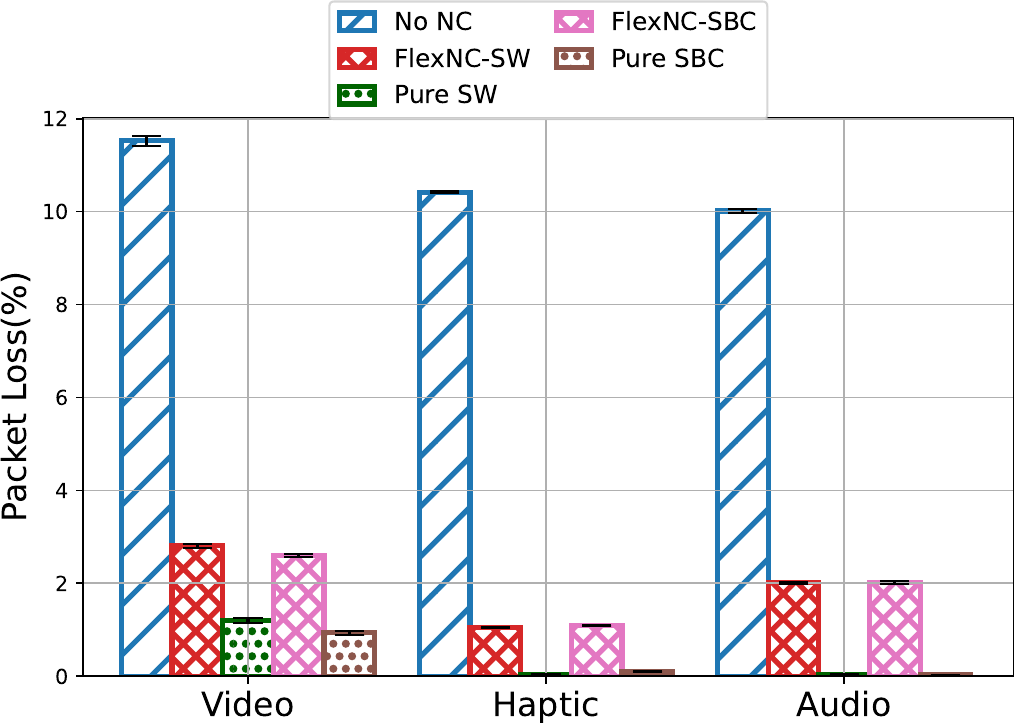}
    \caption{Packet loss measurement under random noise model for different types of traffic (video, haptic, and audio). 
    }
    \vspace{-4mm}
    \label{fig:pkt-loss-rand}
\end{figure}

\paragraph{Packet Loss}
Our goal is to investigate the effectiveness of the proposed FlexNC algorithm to flexibly enhance the 5G System's reliability. To achieve this goal, we examine packet loss in various traffic scenarios. 
Fig.~\ref{fig:pkt-loss-rand} illustrates the packet loss characteristics of the uncoded scenario in comparison to the FlexNC and Pure NC scenarios.
For the No NC scenario, we observe a packet loss of 11.5\%, 10.4\% and 10\% packet loss in video, haptic and audio traffic respectively, which is consistent with the 10\% loss probability set in tc-netem. The video packets are MPEG-2 Transport (MP2T) Streams, which are fragmented to fit over RTP. This MP2T fragmentation can increase packet loss, particularly under limited bandwidth conditions. 
We can further observe from Fig. \ref{fig:pkt-loss-rand} that the packet loss performances of algorithms using SBC and SW schemes for each of the respective traffic types are fairly similar.
Considering the overall configuration for FlexNC with part of the traffic not experiencing any NC, the observed packet losses are well within the expectations of fractional losses.

The packet loss for video traffic in Pure SBC and FlexNC-SBC is 0.9\% and 2.6\%, respectively, while for Pure SW and FlexNC-SW, the video traffic incurs packet loss of 1.2\% and 2.8\%, respectively. 
In the FlexNC scenario, 25\% of the video traffic does not experience any network coding. Thus, 25\% of the video traffic will be experiencing a random loss of 10\% (from tc-netem) while the remaining 75\% of video traffic can recover losses with redundancy. Therefore, theoretically, we should obtain a loss of around 2.5\% for the video traffic. This result is consistent with the video packet loss value for FlexNC-SBC and FlexNC-SW.
10\% of the haptic traffic does not experience any network coding in the FlexNC scenario. This results in 10\% of the haptic traffic experiencing a random loss of 10\% and the remaining 90\% of haptic traffic can recover losses with network coding. Theoretically, we should obtain a loss of around 1\% for the haptic traffic. We obtain similar results in our testbed. We observe packet loss of 1.1\% and 1\% haptic traffic in the FlexNC-SW and FlexNC-SBC scenarios.  

\subsubsection{Bursty Noise}

Next, we present results for the FlexNC algorithm under bursty noise conditions.

\begin{figure*}[t]
    \begin{center}
        \subfigure[]{
        \includegraphics[width=5.5cm]{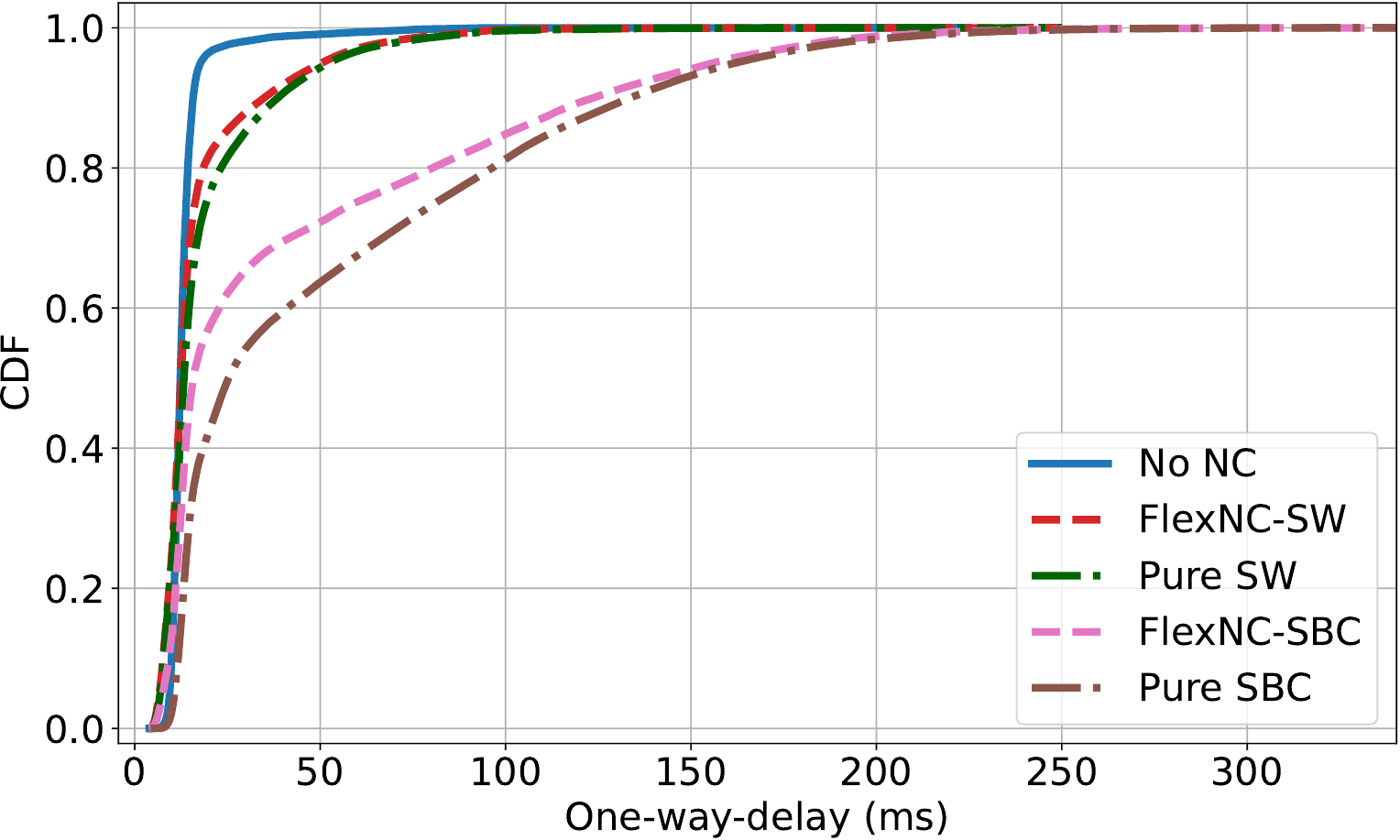} \hfill
        }
        \subfigure[]{
        \includegraphics[width=5.5cm]{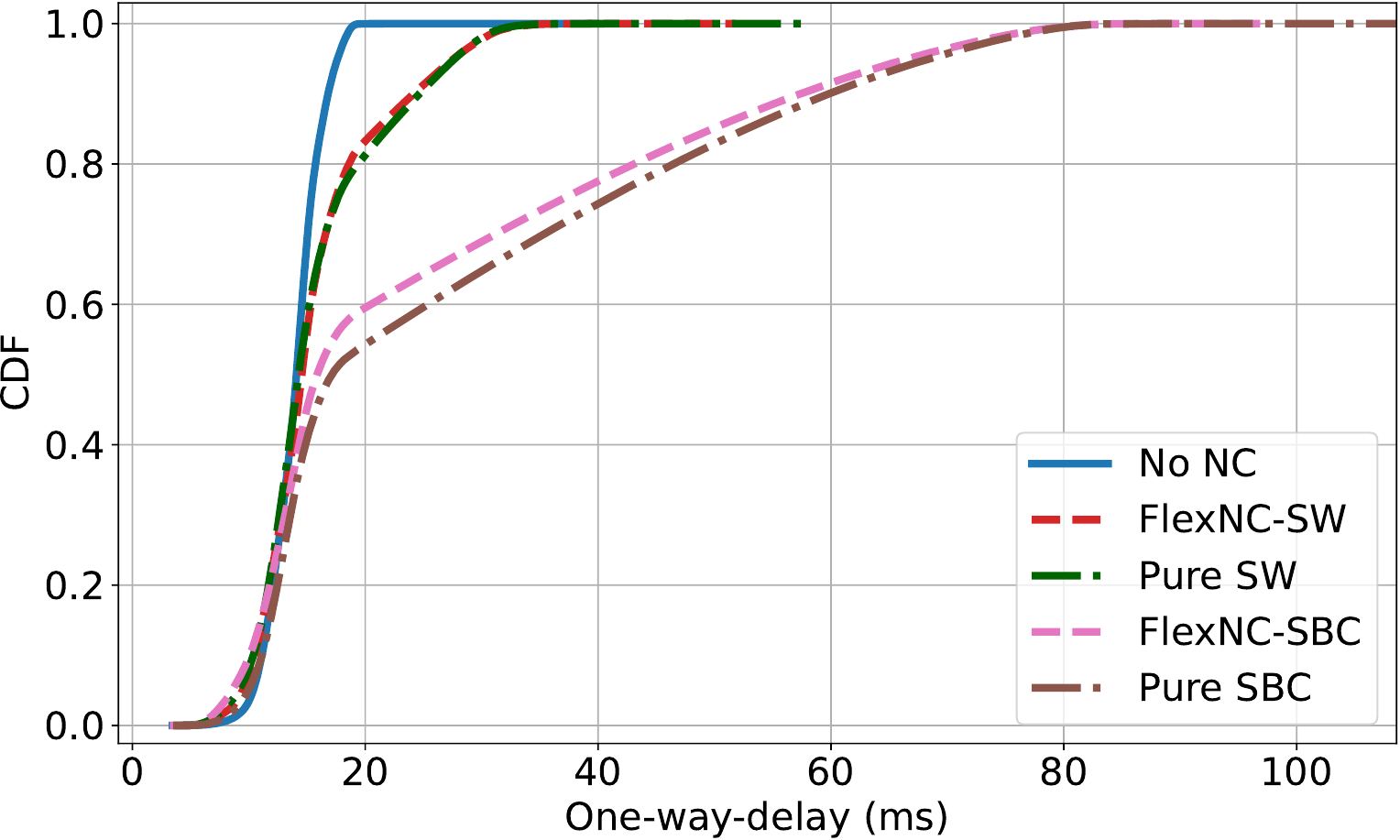} \hfill
        }
        \subfigure[]{
        \includegraphics[width=5.5cm]{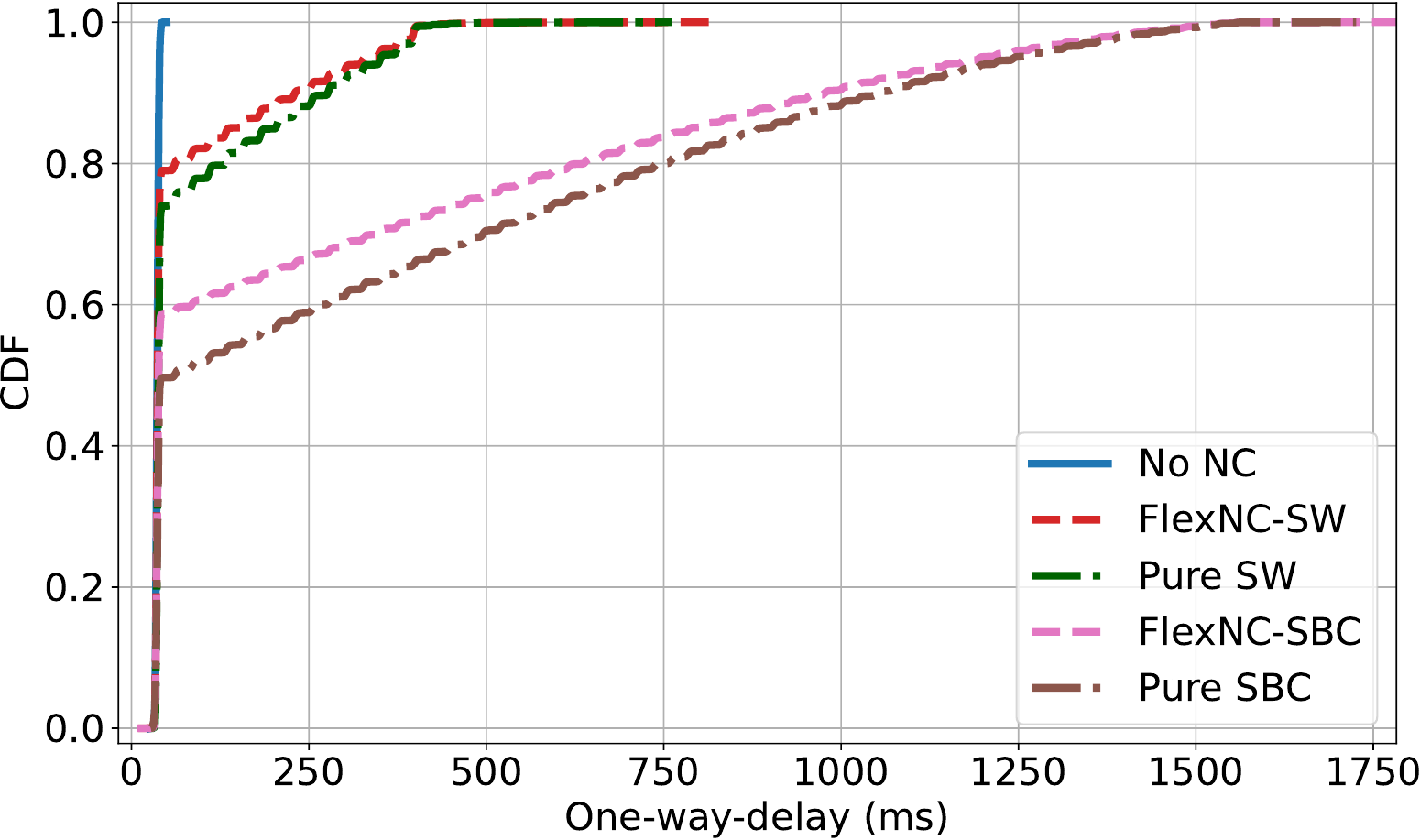} \hfill
        }
    \end{center}
    \caption{Combined overview of OWD measurements in downlink under bursty noise model for all received (a) video, (b) haptic, and (c) audio packets across all test runs.
    }
    \label{fig:owd-burst-TR}
\end{figure*}

\begin{figure*}[t]
    \begin{center}
        \subfigure[]{
        \includegraphics[width=5.5cm]{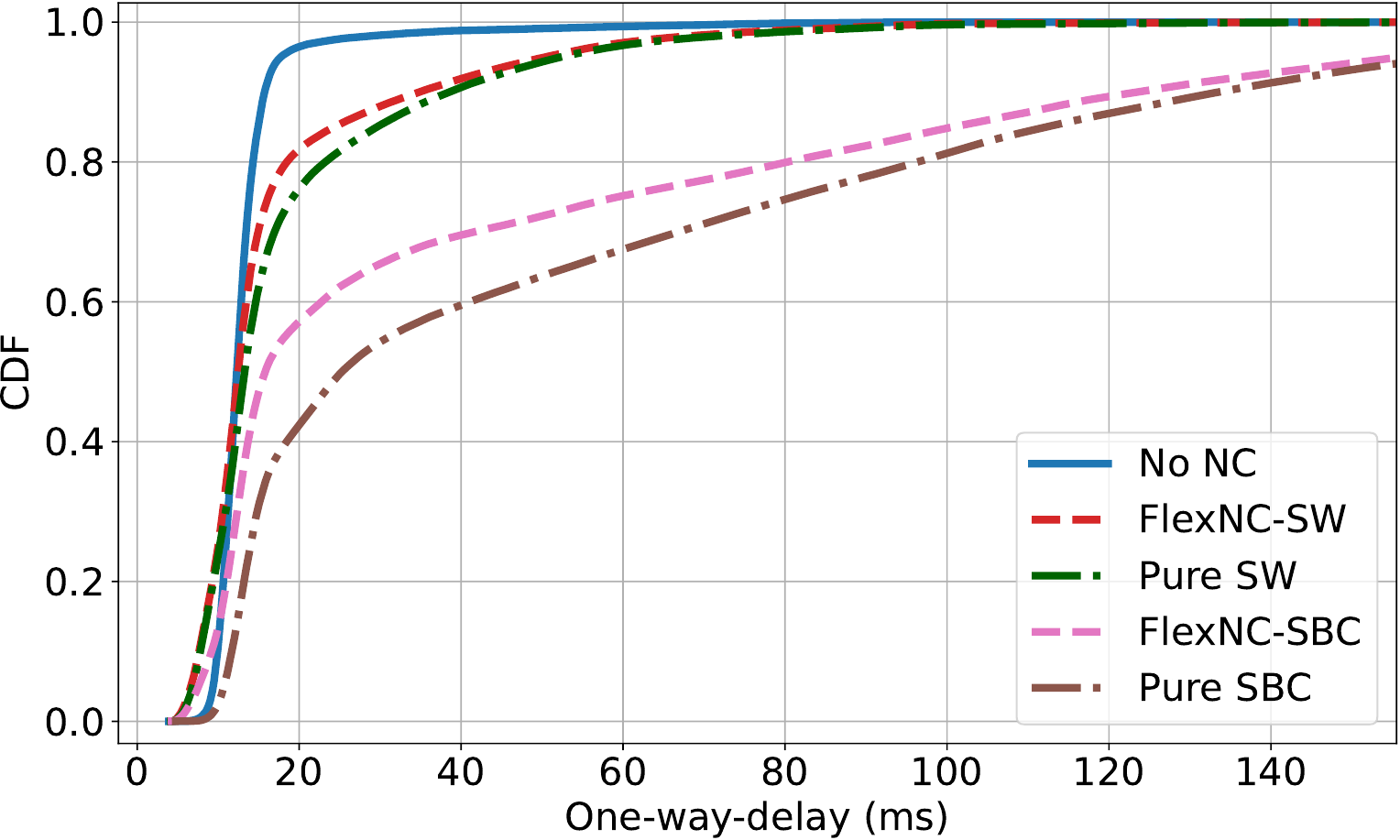} \hfill
        }
        \subfigure[]{
        \includegraphics[width=5.5cm]{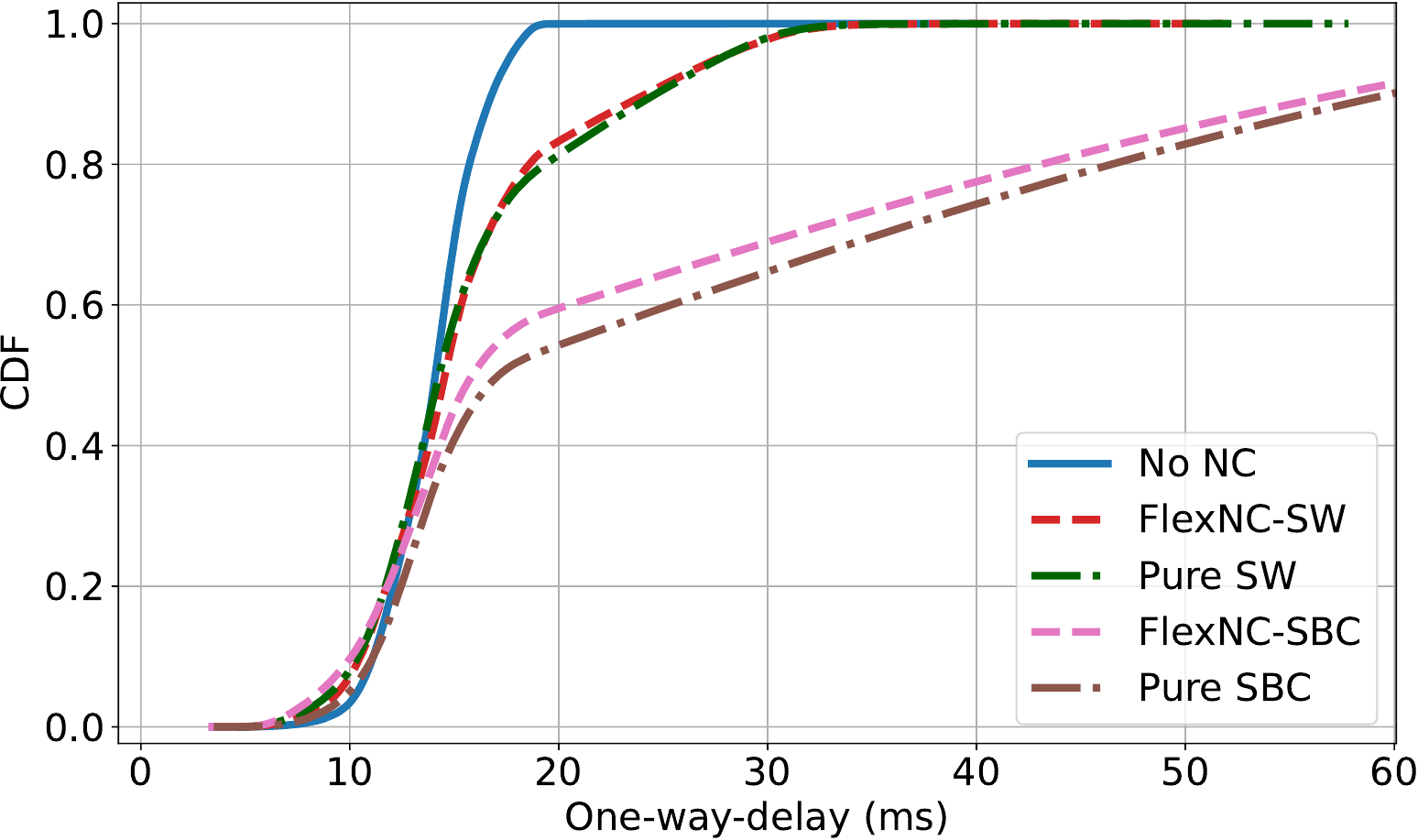} \hfill
        }
        \subfigure[]{
        \includegraphics[width=5.5cm]{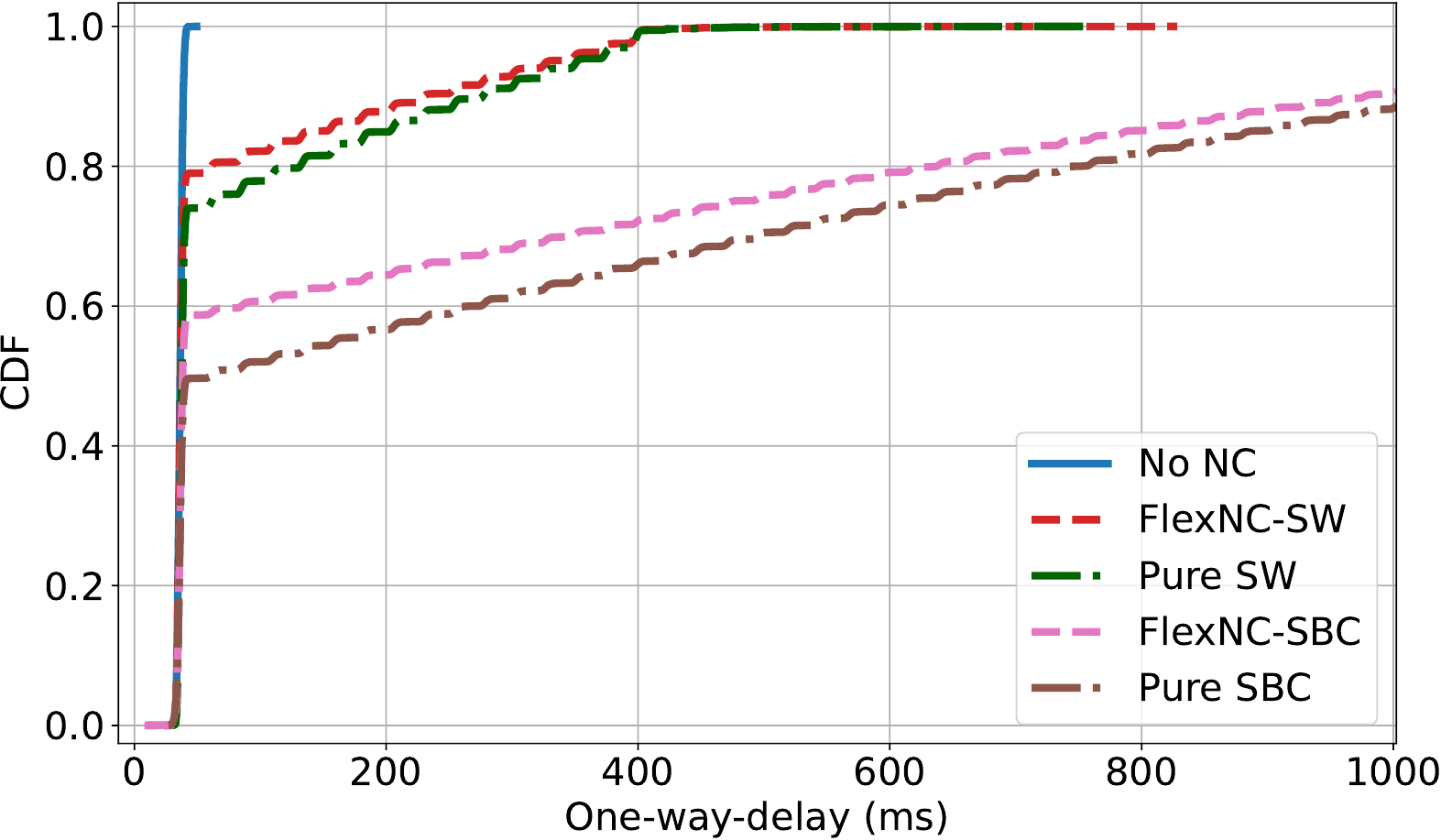} \hfill
        }
    \end{center}
    \caption{Combined overview of OWD measurements in downlink under bursty noise model for all received (a) video, (b) haptic, and (c) audio packets across all test runs, highlighting the region of interest from Fig.~\ref{fig:owd-burst-TR}.}
    \label{fig:owd-burst}
\end{figure*}

\paragraph{Latency / OWD} 
Similar to the random noise scenario, we analyze how much extra delay is needed with the application of network coding on the entire traffic and compare it with FlexNC for the bursty noise model. 
As illustrated in Fig.~\ref{fig:owd-burst-TR}, the OWD for downlink transmissions of the various traffic types is again the lowest for the uncoded scenario. Due to higher waiting times in SBC as in Sec.~\ref{sec:eval-rand-flexnc-owd}, the improvement in the OWD for FlexNC-SBC is more prominent than that of FlexNC-SW from their respective pure network coding versions.

We observed in Fig. \ref{fig:owd-burst-TR}(a) that in FlexNC-SBC, 60\% of the received video packets have OWD value below or equal to 22.8\,ms while for Pure SBC 60\% of the received video packets have OWD value below or equal to 41.1\,ms. Consequently, we observe a reduction of 60\% of the received video packet's OWD by 18.3\,ms with FlexNC-SBC. In FlexNC-SW, 60\% of the received video packets have an OWD value below or equal to 13.4\,ms and for Pure SW the OWD is below or equal to 14.6\,ms. Thus, using FlexNC-SW, 60\% of the received video packets can lower their OWD by 1.2\,ms. In the case of No NC, 60\% of the received video packets have an OWD value below or equal to 12.8\,ms.
We further observe in Fig. \ref{fig:owd-burst-TR}(a) the OWD of almost all received video packets are obtained within 100ms. Meanwhile, almost all received video packets for Pure SW and FlexNC-SW are within 150\,ms. Whereas Pure SBC and FlexNC-SBC can receive most video packets within 300\,ms. 

Next, we observed in Fig. \ref{fig:owd-burst-TR}(b) that the OWD value of 60\% of the received haptic packets for FlexNC-SBC is below or equal to 20.5\,ms while for Pure SBC, the OWD values are below or equal to 25.4\,ms. Therefore, with FlexNC-SBC 60\% of the received haptic packets can reduce their OWD by 4.9\,ms. We observed in FlexNC-SW, 60\% of the received haptic packets have OWD values below or equal to 15.3\,ms and for Pure SW, OWD is below or equal to 15.2\,ms. We notice a small decrease of 0.1\,ms in the OWD of 60\% of the received haptic packets by using FlexNC-SW. Without any network coding, 60\% of the received haptic packets have an OWD value below or equal to 14.5\,ms. We do not consider the outliers observed in four haptic test runs in our evaluation. 
Fig. \ref{fig:owd-flexnc-rand-TR}(b) shows that for the No NC scenario, almost all received haptic packets are obtained within 20ms. Whereas most haptic packets are received by Pure SW and FlexNC-SW in 35\,ms. For Pure SBC and FlexNC-SBC, most haptic packets are received within 45\,ms.

We observed in Fig. \ref{fig:owd-burst-TR}(c) that the OWD value of 60\% of the received audio packets for FlexNC-SBC within 85\,ms while the OWD value for Pure SBC is within 274.4\,ms. Using FlexNC-SBC instead of Pure SBC results in an OWD value of 60\% of the received audio packets being reduced by 189.4\,ms. Meanwhile, for FlexNC-SW, 60\% of the received audio packets' OWD value is below or equal to 37.5\,ms and for Pure SW the OWD values are below or equal to 38.4\,ms. When FlexNC-SW rather than Pure SW is used, 60\% of the received audio packets' OWD are reduced by 50.7\,ms. Finally, we observed that 60\% of the received haptic packets in No NC scenario have OWD value below or equal to 36.9\,ms.
Most received audio packets have OWD within 45\,ms. Due to the high IPD (24\,ms) of audio packets increasing the waiting time at the decoder, the OWD for audio traffic is most affected by the inclusion of network coding (particularly SBC) schemes. Therefore, Pure SBC and FlexNC-SBC need around 1600\,ms to receive most of the audio packet. SW scheme has a smaller waiting time because redundant packets are spread amongst smaller batches of original packets, thus, Pure SW and FlexNC-SW have within 600\,ms for most received audio packets.

\begin{figure}[t]
    \centering
    \includegraphics[width=0.95\columnwidth]{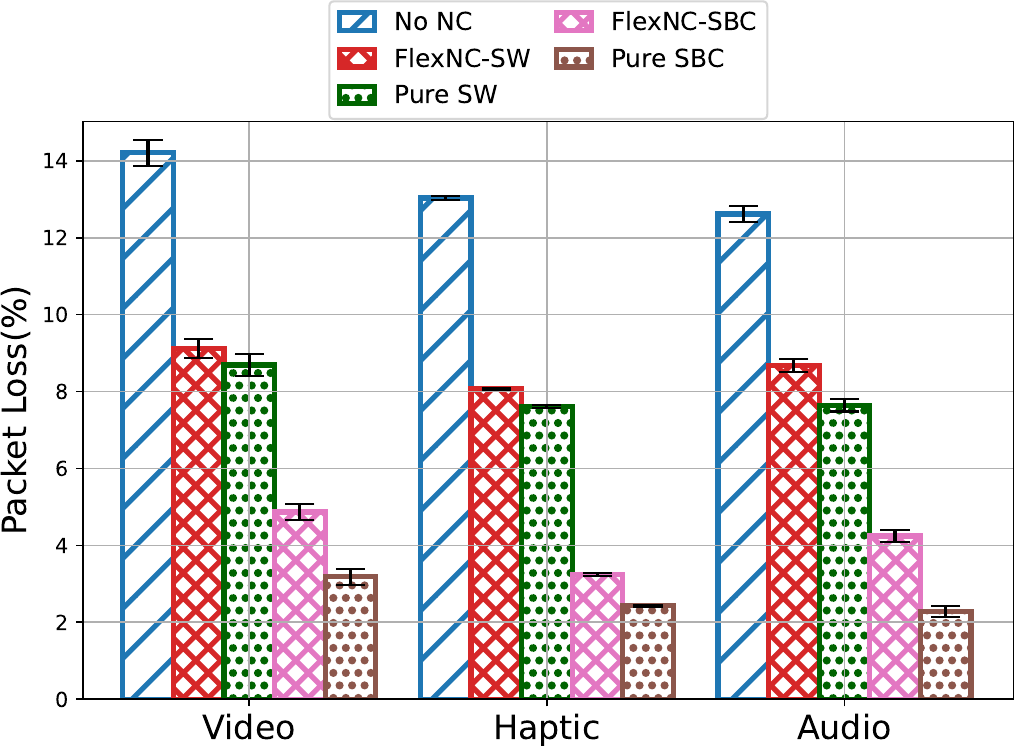}
    \caption{Packet loss measurement under bursty noise model for different types of traffic (video, haptic, and audio).}
    \vspace{-4mm}
    \label{fig:pkt-loss-burst}
\end{figure}

\paragraph{Packet Loss} 
To have a comprehensive evaluation of the proposed FlexNC algorithm we now assess its packet loss performance in bursty environments for different traffic scenarios. 
As illustrated in Fig.~\ref{fig:pkt-loss-burst}, the No NC scenario has a packet loss of 14.2\%, 13\% and 12.6\% for video, haptic, and audio traffic, respectively. 
All NC variants have significantly lower packet loss rates with FlexNC schemes performing slightly worse than their regular counterparts. 
These differences are more pronounced for the SBC variants (about 1--2\%) than the SW variants (about 0.5--1\%).

The packet loss for video traffic in Pure SBC and FlexNC-SBC is 3.2\% and 4.2\% respectively. While for Pure SW and FlexNC-SW, the video traffic incurs packet loss of 8.7\% and 9.1\%. For haptic traffic, we observed a packet loss of 2.4\% and 3.2\% haptic traffic in the Pure SBC and FlexNC-SBC respectively. Whereas for the Pure SW and FlexNC-SW, haptic traffic's packet loss is 7.6\% and 8.1\% respectively. Lastly, we observe the packet loss for audio traffic is 2.3\% and 4.2\% for Pure SBC and FlexNC-SBC respectively. Meanwhile, audio traffic's packet loss for Pure SW and FlexNC-SW is 7.6\% and 8.7\% respectively. 
This is because the SBC scheme does network coding over a whole generation (64) of original packets while SW spreads redundant packets over smaller patches (16) of original packets. 
Bursty noise is characterized by intermittent bursts of high noise levels interspersed with periods of low noise or silence. For example, if the high noise causes 15 original packets to be lost during a particular generation, then SBC can recover them from 49 (64-15) original and 32 coded redundant packets. However, if the same case occurred for SW, then 1 (16-1) original packet and 8 redundant packets cannot recover the lost packets. 

\subsubsection{Summary}
So far, we have observed that FlexNC-SBC is more effective in reducing OWD than FlexNC-SW. And SBC scheme is more effective in reducing packet loss than the SW scheme for burst noise. Therefore, we can demonstrate the benefit of applying FlexNC-SBC in place of Pure NC in burst loss conditions.

\subsection{RecNet}
We next present RecNet evaluation results for latency/OWD and packet losses. 
We aim to analyze the performance of network coding in the 5G System with the inclusion of the recoder. The recoder leverages INC to code packets at the relay node. Due to the recoder being compatible with block coding schemes in the encoder, we evaluate RecNet over only the SBC scheme. Moreover, the switch on which we deploy the recoder does not support the packages for the Linux NetEm server, so we cannot model the bursty noise model in the switch. Therefore, we implement iptables on the decoder to apply packet loss on incoming packets. As Iptables only allows a random loss model, we only evaluate RecNet for random noise conditions.

\subsubsection{One-UE Topology}
With this topology, we evaluate the OWD and packet loss performance in downlink transmission when network coding with the recoder (RecNet) is compared with the case without network coding (No NC) and network coding without the recoder (Pure SBC). In the case of the one-UE topology of the RecNet algorithm, we evaluate video, haptic and audio traffic. 

\begin{figure*}[t]
    \begin{center}
        \subfigure[]{
        \includegraphics[width=8.6cm]{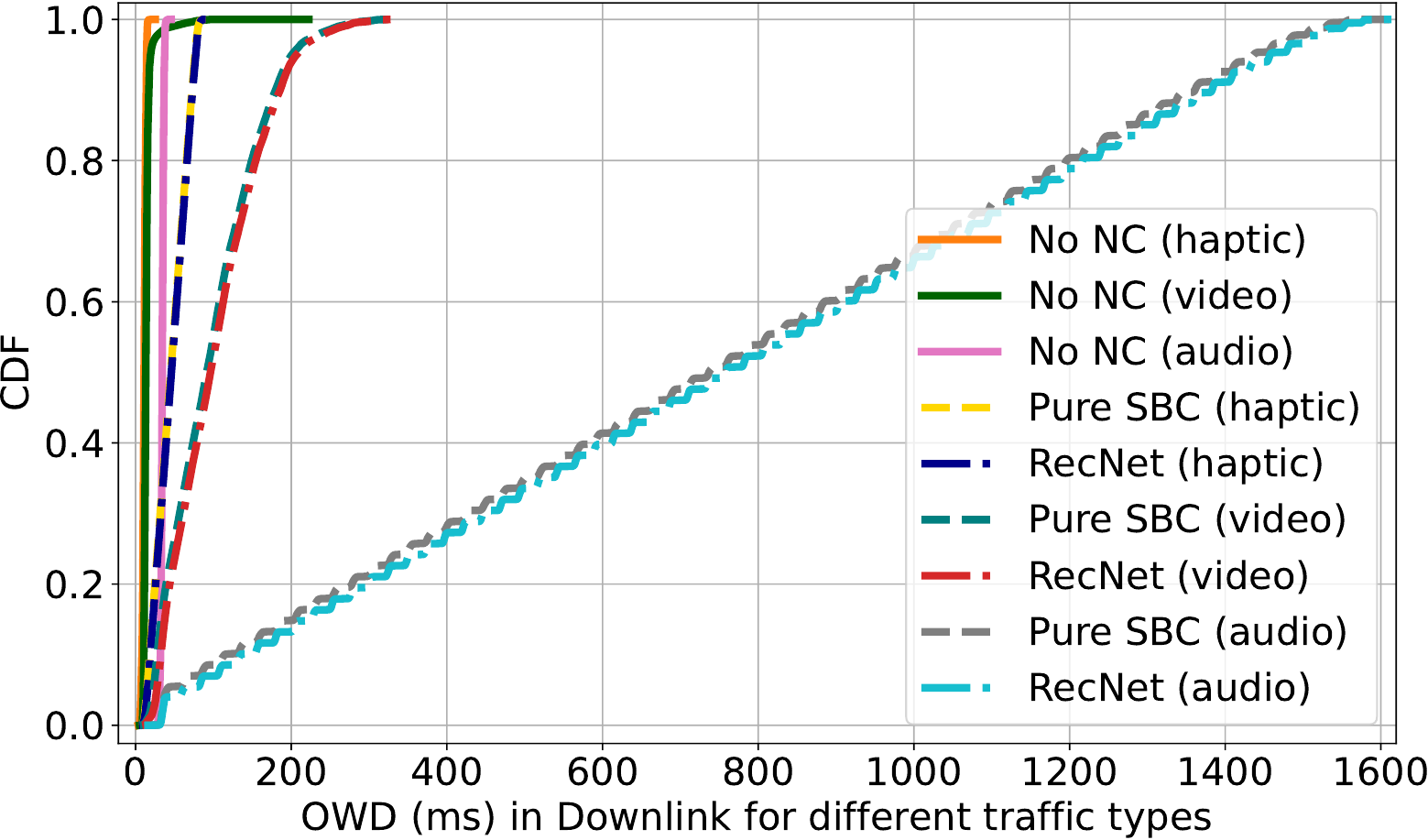} \hfill
        }
        \subfigure[]{
        \includegraphics[width=8.6cm]{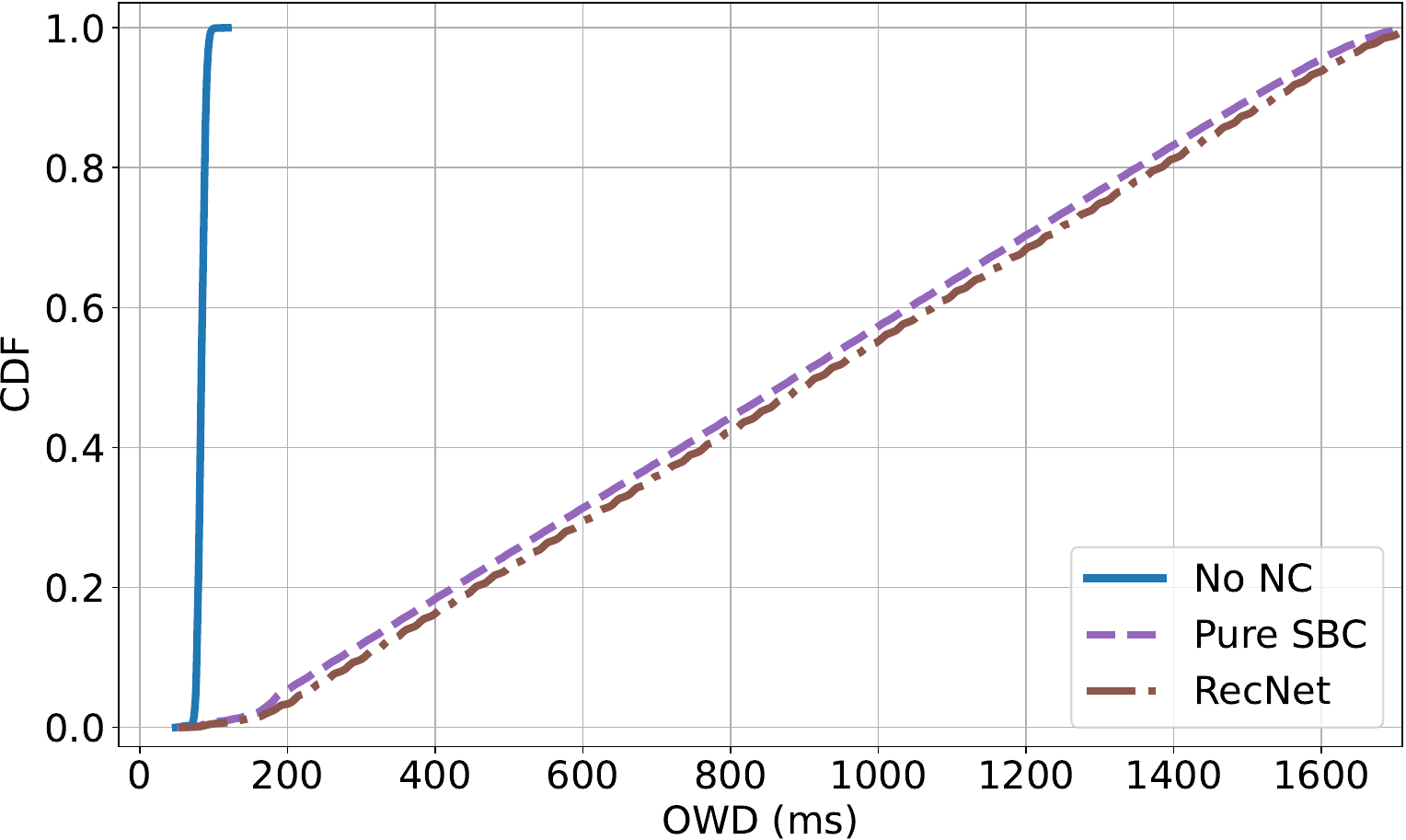} \hfill
        }
    \end{center}
    \caption{Combined overview of OWD measurements under random noise model in (a) one-UE topology (downlink) for all received video, haptic, and audio packets, and (b) two-UE topology for all received audio packets across all test runs.}
    \label{fig:owd-recnet-TR}
\end{figure*}

\begin{figure*}[t]
    \begin{center}
        \subfigure[]{
        \includegraphics[width=8.6cm]{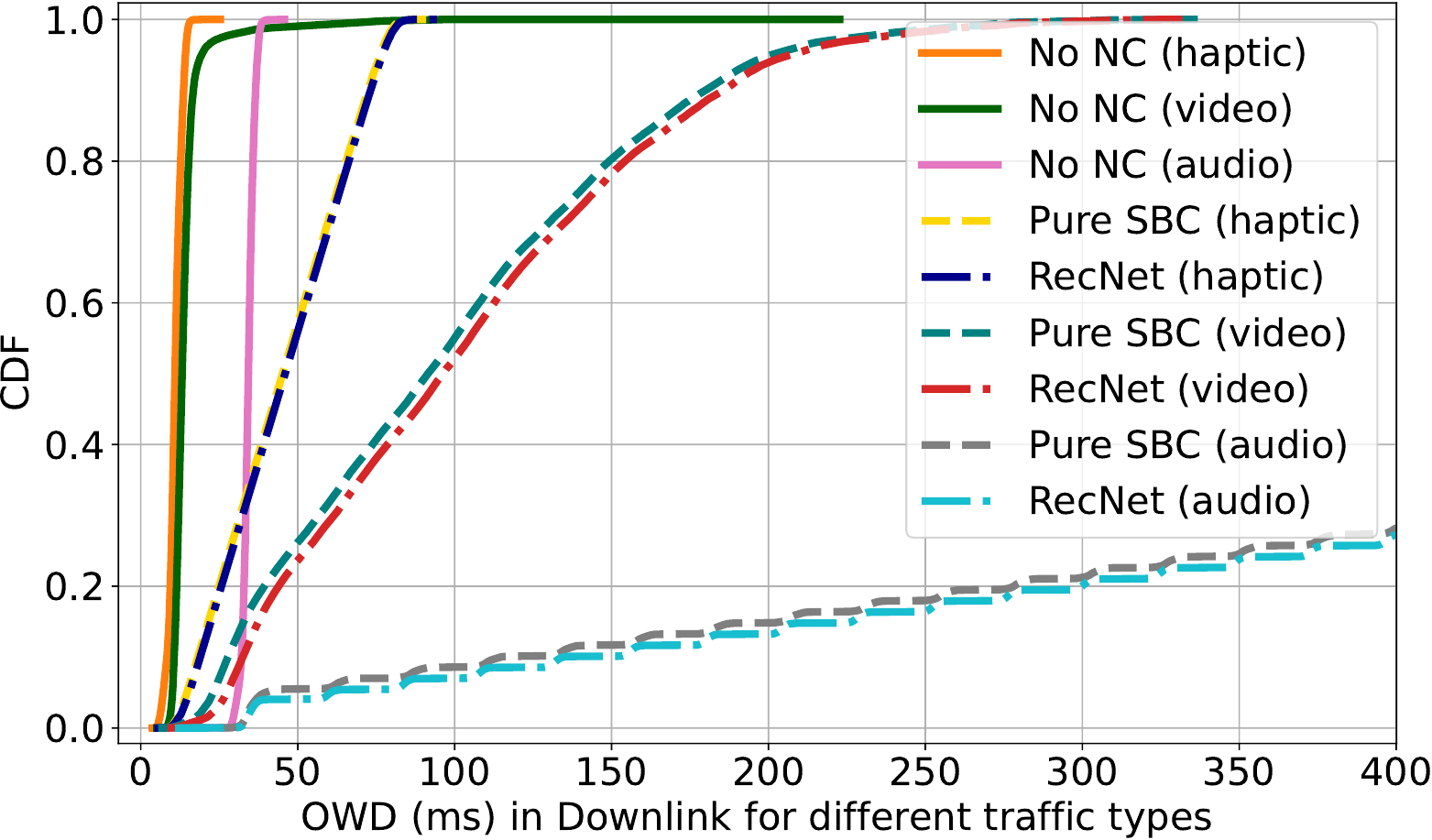} \hfill
        }
        \subfigure[]{
        \includegraphics[width=8.7cm]{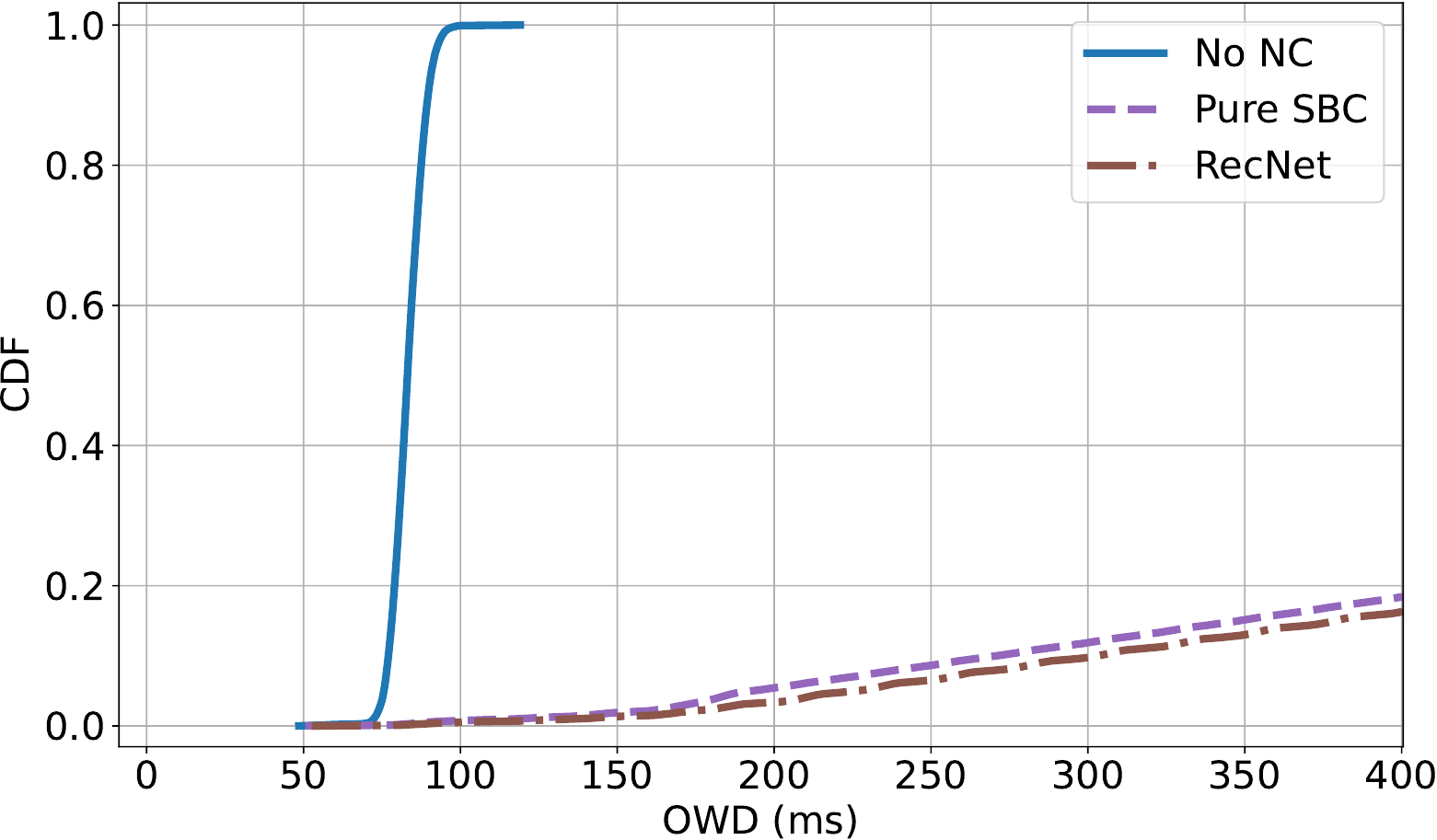} \hfill
        }
    \end{center}
    \caption{Combined overview of OWD measurements under random noise model in (a) one-UE topology (downlink) for all received video, haptic, and audio packets, and (b) two-UE topology for all received audio packets across all test runs, highlighting the region of interest from Fig.~\ref{fig:owd-recnet-TR}.}
    \label{fig:owd-recnet}
\end{figure*}

\paragraph{Latency / OWD} 
We observe from Fig.~\ref{fig:owd-recnet-TR}(a) that the No NC scenarios  have the lowest OWD values compared to the other scenarios. 
We additionally observe that the OWD values for Pure SBC and RecNet for different traffic types are very close to each other. 
60\% of the received video packets have OWD within 107.8\,ms and 112.5\,ms for Pure SBC and RecNet, respectively. This shows that for 60\% of the received video packets 4.7\,ms of additional delay is needed by the recoder. Similarly, 60\% of the received haptic packets have OWD within 51.9\,ms and 52.7\,ms for Pure SBC and RecNet respectively. Thus, 60\% of the received haptic packets consume 0.8\,ms of additional delay for recoding. Lastly, for 60\% of the received audio packets, OWD is within 884.9\,ms and 904.9\,ms for Pure SBC and RecNEt respectively. Consequently, 60\% of the received haptic packets need 20\,ms of additional delay for recoding.
This shows that integrating the recoder in the SBC algorithm does not have much impact on the OWD for downlink transmission of video, haptic and audio traffic in a 5G System.

\begin{figure*}[t]
    \begin{center}
        \subfigure[]{
        \includegraphics[width=8.6cm]{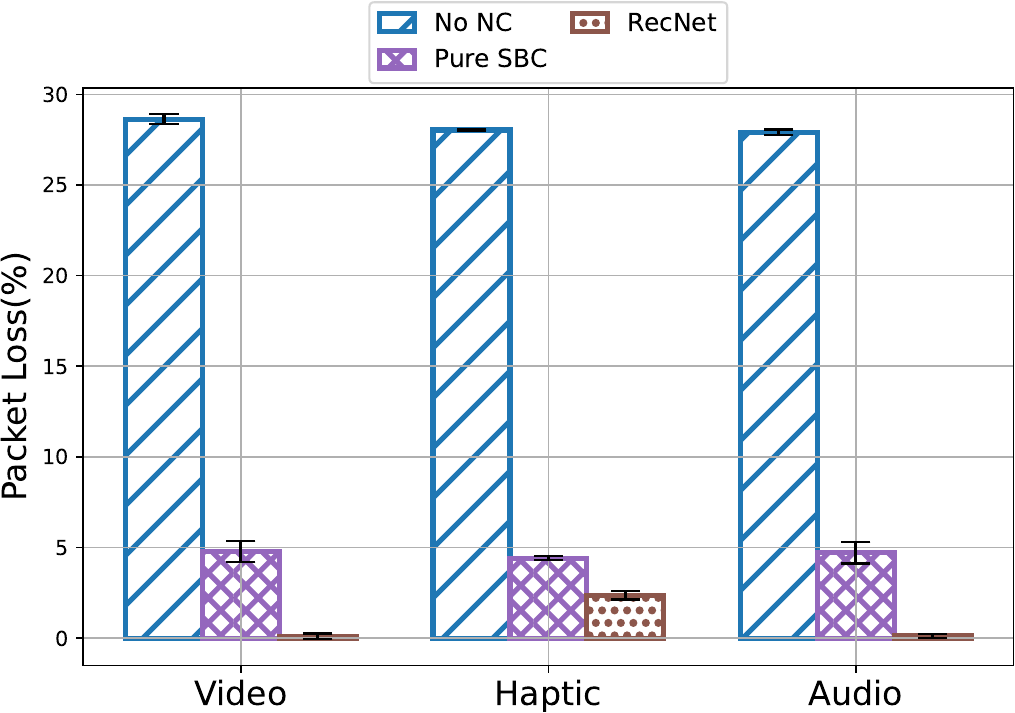} \hfill
        }
         \hspace {1cm}
        \subfigure[]{
        \includegraphics[width=4.9cm]{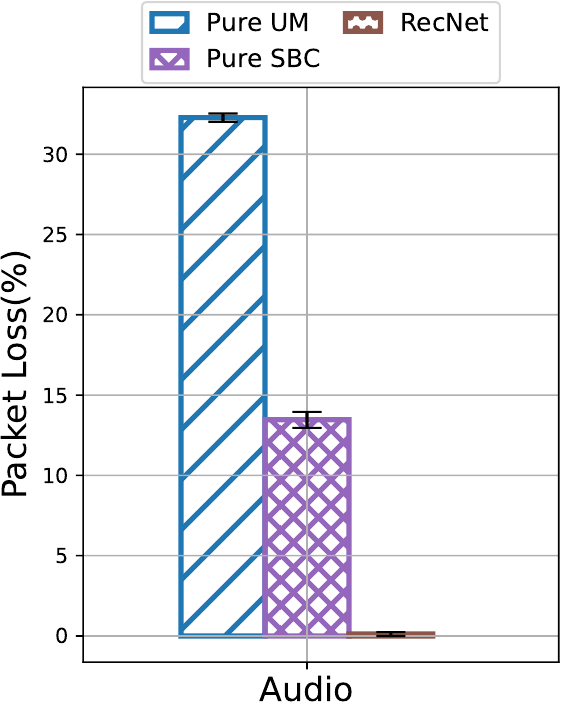} \hfill
        }
    \end{center}
    \caption{Packet loss measurement under random noise model in (a) one-UE topology (downlink) for different types of traffic (video, haptic, and audio), and (b) two-UE topology for audio traffic.}
    \label{fig:loss-recnet}
\end{figure*}

\paragraph{Packet Loss} To emulate the remote UE scenario in a 5G System, we apply random packet loss of 10\% for incoming packets at the relay node and 20\% for incoming packets at the destination node (decoder). In the No NC scenario, video, haptic and audio traffic incur packet loss of 28.6\%, 28\%, and 27.9\%, respectively. 
When an encoder and a decoder are used, but the relay node forwards the packet without recoding, the packet loss for video, haptic and audio traffic drops to 4.8\%, 4.4\%, and 4.7\%, respectively. However, with RecNet the packet loss for video, haptic and audio traffic is further reduced to 0.1\%, 2.4\%, and 0.1\%, respectively, as shown in Fig.~\ref{fig:loss-recnet}(a). This reduction is achieved using the recoder without incurring heavy additional delay, as seen in Fig.~\ref{fig:owd-recnet-TR}(a).

\subsubsection{Two-UE Topology}
The two-UE topology involves both uplink and downlink communication. Due to the two wireless 5G transmissions, we obtain unstable results for traffic with low IPD values, such as video and haptic traffic, leading to queue build-up on both uplink and downlink. Specifically, we observed heavy packet loss for video traffic even without applying a loss model, due to a combination of high payload size and low IPD of video packets. Therefore, in this topology, we only evaluate the audio traffic, which has a higher IPD and moderate payload size. 

\paragraph{Latency / OWD}
Fig.~\ref{fig:owd-recnet-TR}(b) shows the CDF plot of OWD of audio packets. As expected the No NC scenario has the lowest OWD, with most received audio packets having an OWD value within 100\,ms. The SBC scheme results in higher OWD due to the long recovery time at the decoder, as discussed extensively in Sec.~\ref{sec:eval-flex-nc}. 

It is evident from Fig.~\ref{fig:owd-recnet-TR}(b) that adding the recoder to Pure SBC does not necessitate a significant amount of extra delay.   
60\% of received audio packets for Pure SBC and RecNet have OWD within 1041.4\,ms and 1073.8\,ms. This shows that 32.4\,ms extra delay was incurred for 60\% of received audio packets by using the recoder. 
Most of the received audio packets have OWD value within 1720\,ms for RecNet and Pure SBC. 
Most importantly, the OWD results of RecNet do not show a significant penalty when compared to Pure SBC. 

\paragraph{Packet Loss} 
For this topology, we apply a random packet loss of 15\% for incoming packets at the relay node and 20\% for incoming packets at the decoder. In the No NC scenario, audio traffic incurs packet loss of 32.3\%. As shown in Fig. \ref{fig:loss-recnet}(b), for Pure SBC (no recoder), the packet loss for audio traffic is 13.5\%. However, RecNet significantly reduces the packet loss for audio traffic to 0.1\%. By employing the recoder, we can reduce packet loss without considerable additional latency as observed in Fig. \ref{fig:owd-recnet}(b).

\subsubsection{Summary}
We have observed that integrating a recoder at the relay can significantly decrease packet loss for remote UEs in both uplink and downlink communication, without incurring much additional latency compared to Pure SBC. Our 1-UE topology test demonstrated the effectiveness of RecNet across different types of traffic.

\section{Conclusion}

In this work, we integrate Random Linear Network Coding (RLNC) into a cloud-native 5G System using the OpenAirInterface Platform, emulating a real-world system. Our approach leverages programmable network coding to address the diverse needs of various applications, leading to the development of two novel algorithms: FlexNC and RecNet.

FlexNC demonstrates a greater delay reduction with SBC nodes than with SW nodes. FlexNC-SBC proves more effective in balancing relatively lower delay with lower packet loss in environments with bursty noise, whereas FlexNC-SW is more effective in achieving these results under random noise conditions, specifically for audio and video traffic.
RecNet demonstrates that recoding on a real switch incurred minimal additional processing delay while significantly decreasing packet loss for remote UEs.
FlexNC and RecNet together can provide a robust framework for enhancing 5G network performance through intelligent traffic management and network coding. These innovations unlock the potential for a more reliable and efficient 5G system, capable of supporting a wide range of applications with varying performance requirements.

Our work identifies different avenues that are worth investigating further. While we have evaluated our proposed algorithms on various time-sensitive traffic types, there is a potential to additionally explore RecNet's performance across a broader mix of traffic types and under different hardware conditions. Incorporating a full hardware implementation, including the air interface, will further advance our work to pave the way for RLNC integration in a real, physical 5G testbed.

\section*{Acknowledgment}
This work was supported by the German Research Foundation 
(DFG, Deutsche Forschungsgemeinschaft) as part of Germany's
Excellence Strategy—EXC 2050/1—Cluster of Excellence ``Centre for Tactile 
Internet with Human-in-the-Loop'' (CeTI) of Technische
Universität Dresden under Project ID 390696704 and the Federal Ministry of Education and Research of Germany in the programme of "Souverän. Digital. Vernetzt." Joint project 6G-life, project identification number: 16KISK001K and the French-German project 5G-OPERA which is funded by the German Federal Ministry of Economic Affairs and Climate Action (BMWK), grant 01MJ22008A.

\bibliographystyle{IEEEtran}
\bibliography{bibliography}

\end{document}